\documentclass{sig-alternate-10pt}

\usepackage{graphicx}
\usepackage{balance}  

\usepackage{ifpdf}
\ifpdf
\else
\fi

\usepackage[T1]{fontenc}
\usepackage{amsmath}
\usepackage{amssymb}
\usepackage{mathrsfs} 
\usepackage{txfonts} 
\usepackage[scaled=.89]{helvet} 
\usepackage{paralist}

\newcommand{\ourtitle}{Query Containment for Highly Expressive Datalog Fragments}

\usepackage{url}
\usepackage[bookmarks=false,pdfpagelabels,bookmarksnumbered,bookmarksopen,bookmarksopenlevel=1,colorlinks=true,
citecolor=black, filecolor=black, linkcolor=black, menucolor=black,
urlcolor=black, pdftitle={\ourtitle}, pdfauthor={Authors},
pdfkeywords={Datalog, first-order rewriting, monadic second-order
logic, query containment}]{hyperref} 

\usepackage{xspace}
\usepackage{rotating}

\newcommand{\containmentMembershipStatement}[3]{Containment of #1 queries in #2 queries can be decided in #3.\xspace}
\newcommand{\containmentHardnessStatement}[3]{Deciding containment of #1 queries in #2 queries is hard for #3.\xspace}

\usepackage{color}

\newtheorem{definition}{Definition}
\newtheorem{example}{Example}
\newtheorem{lemma}{Lemma}
\newtheorem{theorem}[lemma]{Theorem}

\newtheorem{proposition}[lemma]{Proposition}

 \usepackage{refcountright}
\newtheorem{innerreptheorem}{Theorem}
\newenvironment{reptheorem}[1]%
{\setcounterref{innerreptheorem}{#1}\addtocounter{innerreptheorem}{-1}%
\begin{innerreptheorem}}
{\end{innerreptheorem}}

\newtheorem{innerreplemma}{Lemma}
\newenvironment{replemma}[1]%
{\setcounterref{innerreplemma}{#1}\addtocounter{innerreplemma}{-1}%
\begin{innerreplemma}}
{\end{innerreplemma}}

\newtheorem{innerrepproposition}{Proposition}
\newenvironment{repproposition}[1]%
{\setcounterref{innerrepproposition}{#1}\addtocounter{innerrepproposition}{-1}%
\begin{innerrepproposition}}
{\end{innerrepproposition}}

\newcommand{\qlang}[1]{{\sf #1}} 
\newcommand{\linq}[1]{\qlang{Lin}{#1}} 
\newcommand{\knestedq}[2]{#1$^{#2}$} 
\newcommand{\nestedq}[1]{\knestedq{#1}{+}} 
\newcommand{\cq}{\qlang{CQ}\xspace}
\newcommand{\ucq}{\qlang{UCQ}\xspace}

\newcommand{\ctworpq}{\qlang{C2RPQ}\xspace}

\newcommand{\nctworpq}[1]{\nestedq{\qlang{C2RPQ}}\xspace}
\newcommand{\datalog}{\qlang{Dlog}\xspace}
\newcommand{\mdatalogconst}{\qlang{MDlog}\xspace} 
\newcommand{\gdatalog}{\qlang{GDlog}\xspace} 
\newcommand{\lindatalog}{\linq{\datalog}\xspace} 
\newcommand{\linmdatalogconst}{\linq{\mdatalogconst}\xspace} 
\newcommand{\lingdatalog}{\linq{\gdatalog}\xspace} 
\newcommand{\mqlang}{\qlang{MQ}}
\newcommand{\mq}{\mqlang\xspace}
\newcommand{\kmq}[1]{\knestedq{\mqlang}{#1}\xspace}
\newcommand{\nmq}{\nestedq{\mqlang}\xspace}
\newcommand{\linmqlang}{\linq{\qlang{MQ}}}
\newcommand{\linmq}{\linmqlang\xspace}
\newcommand{\klinmq}[1]{\knestedq{\linmqlang}{#1}\xspace}
\newcommand{\nlinmq}{\nestedq{\linmqlang}\xspace}
\newcommand{\gqlang}{\qlang{GQ}}
\newcommand{\gq}{\gqlang\xspace}
\newcommand{\kgq}[1]{\knestedq{\gqlang}{#1}\xspace}
\newcommand{\ngq}{\nestedq{\gqlang}\xspace}
\newcommand{\lingqlang}{\linq{\qlang{GQ}}}
\newcommand{\lingq}{\lingqlang\xspace}
\newcommand{\klingq}[1]{\knestedq{\lingqlang}{#1}\xspace}
\newcommand{\nlingq}{\nestedq{\lingqlang}\xspace}

\newcommand{\complclass}[1]{{\sc #1}\xspace} 

\newcommand{\ACzero}{\complclass{AC$_0$}}
\newcommand{\NLogSpace}{\complclass{NLogSpace}}
\newcommand{\PTime}{\complclass{P}}
\newcommand{\NP}{\complclass{NP}}

\newcommand{\PH}{\complclass{PH}}
\newcommand{\PSpace}{\complclass{PSpace}}
\newcommand{\NPSpace}{\complclass{NPSpace}}
\newcommand{\ExpTime}{\complclass{ExpTime}}

\newcommand{\ExpSpace}{\complclass{ExpSpace}}
\newcommand{\TwoExpTime}{\complclass{2ExpTime}}

\newcommand{\ThreeExpTime}{\complclass{3ExpTime}}
\newcommand{\kExpTime}[1]{\complclass{#1ExpTime}}
\newcommand{\kExpSpace}[1]{\complclass{#1ExpSpace}}

\newcommand{\tuple}[1]{\langle{#1}\rangle}
\newcommand{\defeq}{\coloneqq}
\newcommand{\signature}{\mathscr{S}}
\newcommand{\arity}{\text{\sf{ar}}} 
\renewcommand{\vec}[1]{\boldsymbol{#1}}

\newcommand{\Inter}{\mathcal{I}} 
\newcommand{\domain}[1]{\mathsf{dom}(#1)} 

\newcommand{\flagconst}{\lambda}
\newcommand{\checkpred}{\mathsf{hit}}
\newcommand{\colorpred}{\mathtt{U}}
\newcommand{\arule}{\rho} 
\newcommand{\aprogram}{\mathbb{P}} 
\newcommand{\aautomaton}{\mathcal{A}} 

\newcommand{\lang}[1]{\ensuremath{\mathbf{#1}}} 
\newcommand{\Ilang}{\lang{C}} 
\newcommand{\Plang}{\lang{P}} 
\newcommand{\Vlang}{\lang{V}} 
\newcommand{\Var}{\mathsf{Var}}

\newcommand{\trees}{\mathrm{Trees}}
\newcommand{\nodes}{\mathrm{Nodes}}
\newcommand{\sq}[1]{\vec{#1}}

\newcommand{\TRpaper}[2]{#2}

\begin{document}

\conferenceinfo{PODS'14,}{June 22--27, 2014, Snowbird, Utah, USA.}
\CopyrightYear{2014}
\clubpenalty=10000 \widowpenalty=10000

\title{\ourtitle{} \TRpaper{(Extended Technical Report)}{}}


\numberofauthors{4}

\author{
Pierre Bourhis\\
       \affaddr{CNRS LIFL University of Lille 1}\\
       \affaddr{$\&$ INRIA Lille Nord Europe}\\
    \affaddr{Lille, FR}
\alignauthor
Markus Kr\"{o}tzsch\\
       \affaddr{Fakult\"{a}t Informatik}\\
       \affaddr{Technische Universit\"{a}t Dresden, DE}\\
\alignauthor
Sebastian Rudolph\\
       \affaddr{Fakult\"{a}t Informatik}\\
       \affaddr{Technische Universit\"{a}t Dresden, DE}\\
}

\maketitle

\begin{abstract}
%
The containment problem of Datalog queries is well known to be undecidable.
There are, however, several Datalog fragments for which containment is
known to be decidable, most notably monadic Datalog and several ``regular'' query languages
on graphs.
Monadically Defined Queries (MQs) have been introduced recently as a joint generalization of
these query languages.

In this paper, we study a wide range of Datalog fragments with decidable query containment
and determine exact complexity results for this problem.
We generalize MQs to (Frontier-)Guarded Queries (GQs), and show that the containment problem is
\ThreeExpTime-complete in either case, even if we allow arbitrary Datalog in the sub-query.
If we focus on graph query languages, i.e., fragments of linear Datalog, then this complexity
is reduced to \kExpSpace{2}.
We also consider nested queries, which gain further expressivity by using predicates that are defined by inner queries.
We show that nesting leads to an exponentially increasing hierarchy for the complexity of query containment,
both in the linear and in the general case.
Our results settle open problems for (nested) MQs, and they paint a comprehensive picture of the state of
the art in Datalog query containment.
\end{abstract}
\section{Introduction}\label{sec:intro}

Query languages and their mutual relationships are a central topic in database research
and a continued focus of intensive study.
It has long been known that first-order logic expressions over the database relations
(represented by \emph{extensional database predicates}, EDBs)
lack the expressive power needed in many scenarios.
Higher-order query languages have thus been introduced, which
allow for the recursive definition of new predicates (so called
\emph{intensional database predicates}, IDBs). Most notably,
Datalog has been widely studied as a very expressive
query language with tractable query answering (w.r.t.\ the size
of the database).

On the other hand, Datalog has been shown to be too expressive a
language for certain tasks which are of crucial importance in database
management. In particular, the \emph{query containment problem} that,
given two queries $Q_1$ and $Q_2$, asks if every answer to $Q_1$ is
an answer to $Q_2$ in every possible database, is undecidable for
full Datalog \cite{Shm87}. However, checking query containment is an
essential task facilitating query optimization, information
integration and exchange, as well as database integrity checking. It
comes handy for utilizing databases with materialized views and, as
part of an offline preprocessing technique, and it may help accelerating
online query answering.

This motivates the question for Datalog fragments that are still
expressive enough to satisfy their purposes but exhibit decidable
query containment. Moreover, once decidability is established, the
precise complexity of deciding containment provides further insights.
The pursuit of these issues has led to a productive and well-established line of research in
database theory, which has already produced numerous results for a
variety of Datalog fragments.

\begin{figure*}
\graphicspath{{figures/}}
\def\svgwidth{13cm}
\mbox{}\hfill\scalebox{0.9}{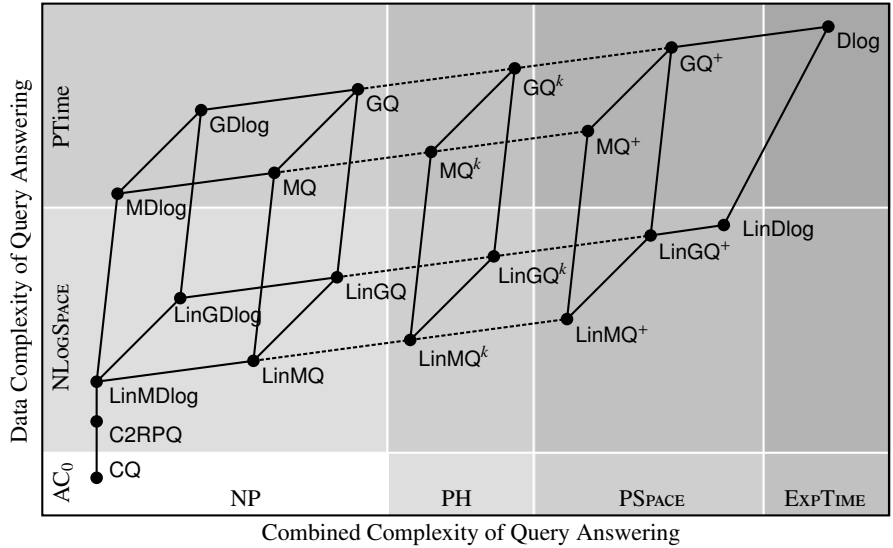}\hfill\mbox{}
\caption{Query languages and complexities; languages higher up in the graph are more expressive\label{fig_querycompl}}
\end{figure*}

\paragraph*{Non-recursive Datalog and unions of conjunctive queries} A
non-recursive Datalog program does not have any (direct or indirect)
recursion and it is equivalent to a union of conjunctive queries
(\ucq) (and thus expressible in first-order logic). The problem of
containment of a Datalog program (in the following referred to as
\datalog) in a union of conjunctive queries is
{\TwoExpTime}-complete \cite{ChaudhuriV97}. Due to the succintness
of non-recursive Datalog compared to \ucq{}s, the problem of
containment of \datalog in non-recursive Datalog is
\ThreeExpTime-complete \cite{ChaudhuriV97}. Some restrictions for
decreasing the complexity of these problems have been considered.
Containment of
\emph{linear} Datalog programs (\lindatalog), i.e., one where rule
bodies contain at most one IDB in a \ucq, is \ExpSpace-complete;
complexity further decreases to \PSpace when the linear Datalog program
is monadic (\linmdatalogconst, see below) \cite{ChaudhuriV94,ChaudhuriV97}.

The techniques to prove the upper bounds in these results are based
on the reduction to the problem of containment of tree automata for
the general case, and to the containment of word automata in the
linear case.


\paragraph*{Monadic Datalog} A monadic Datalog (\mdatalogconst) program is
a program containing only unary intensional predicates.
The problem of containment for \mdatalogconst is \TwoExpTime
complete. The upper bound is well known since the 80's
\cite{Cosmadakis88}, while the lower bound has been established only recently
\cite{BenediktBS12}. Finally, the containment of \datalog in a
monadic \mdatalogconst is also decidable. It is a straightforward
application of Theorem~5.5 of \cite{Courcelle91}.\footnote{We thank Michael Benedikt for this observation.}
So far, however, tight bounds have not been known for this result.

\paragraph*{Guarded Datalog} Guarded Datalog (\gdatalog) allows
the use of intensional predicates with unrestricted arities, however
for each rule, the variables of the head should appear in a single
extensional atom appearing in the body of the rule.
While this notion of (frontier-)guarded rules is known for a while \cite{CaliGK08,BLMS11:decline}, the first
use of \gdatalog as a query language seems to be only recent \cite{BaranyCO12}.
\gdatalog is a proper extension of \mdatalogconst, since monadic
rules can always be rewritten into guarded rules \cite{BaranyCO12}.
It is know that query containment for \gdatalog is \TwoExpTime-complete,
a result based on the decidability of the
satisfiability of the guarded negation fixed point logic \cite{BaranyCS11}.
%
%

\paragraph*{Navigational Queries} Conjunctive two-way regular path queries
(\ctworpq{}s) generalize conjunctive queries (\cq{}s) by regular
expressions over binary predicates
\cite{regularpathqueries1,regularpathqueries2}. Variants of this
type of queries are used, e.g., by the XPath query language for
querying semi-structured XML data. Recent versions of the SPARQL~1.1
query language for RDF also support some of regular expressions that
can be evaluated under a similar semantics. Intuitively, \ctworpq is
a conjunct of atoms of the form $xLy$ where $L$ is a two-way regular
expression. A pair of nodes $\tuple{n_1,n_2}$ is a valuation of the pair
$\tuple{x,y}$ if and only if there exists a path between $n_1$ and $n_2$ matching
$L$. The containment of queries in this language was shown to be
\ExpSpace-complete
\cite{regularpathqueries1,CalvaneseGLV03,AbiteboulV99,Deutsch2001}.
The containment of \datalog in \ctworpq is
\TwoExpTime-complete \cite{CalvaneseGV05}.

\paragraph*{Monadically Defined Queries} More recently, Monadically Defined
Queries (\mq{}s) and their nested version (\nmq{}s) have been
introduced \cite{RK13:flagcheck} as a proper generalization of
\mdatalogconst which also captures (unions of) \ctworpq{}s. At the
same time, they are conveniently expressible both in \datalog and
monadic second-order logic. Yet, as opposed to these two, \mq{}s and
\nmq{}s have been shown to have a decidable containment
problem, but no tight bounds were known so far.

\medskip
In spite of these continued efforts, the complexity of query containment
is still unclear for many well-known Datalog fragments, especially for the
most expressive ones.
In this paper, we thus study a variety of known and new query languages in more detail.
Figure~\ref{fig_querycompl} gives an overview of
all Datalog fragments we consider, together with their respective
query-answering complexities.

We provide a detailed complexity analysis of the
mutual containment between queries of the aforementioned (and some
new) formalisms.
This analysis is fine-grained in the sense that---in
the case of query formalisms that allow for nesting---precise
complexities depending on the nesting depth are presented. Moreover,
we consider the case where the used rules are restricted to linear
Datalog.

\begin{itemize}

\item
We introduce \emph{guarded queries} (\gq{}s) and their nested
versions (\ngq{}s), Datalog fragments that properly generalize
\mq{}s and \nmq{}s, respectively, while featuring the same data and
combined complexities for query answering. On
the other hand, already unnested \gq{}s subsume \gdatalog. We also
consider the restrictions of all these queries to the linear Datalog
case and observe that this drops data complexities to \NLogSpace
whereas it does not affect combined complexities.
\item
By means of sophisticated automata-based techniques involving
iterated transformations on alternating two-way automata, we show a
generic upper bound stating that containment of \datalog in nested
guarded queries of depth $k$ (\kgq{k}) can be decided in
\kExpTime{$(k+2)$}. Additionally we show that going down to
\gdatalog on the containment's right-hand side allows deciding it in
\TwoExpTime.
\item
Inductively defining alternating Turing machine simulations on tapes
of $(k+1)$-exponential size, we provide a matching generic lower
bound by showing that containment of \mdatalogconst in \kmq{k} is
\kExpTime{$(k+2)$}-hard. Together with the upper bound, this
provides precise complexities for all cases, where the left-hand
side of the containment is any fragment between \mdatalogconst and
\datalog (cf. Fig.~\ref{fig_querycompl}) and the right-hand side is
any of \mq, \gq, \kmq{k}, \kgq{k}, \nmq, \ngq. In particular, this
solves the respective open questions from \cite{RK13:flagcheck}: \mq
containment is \kExpTime{$3$}-complete and \nmq containment is
\complclass{NonElementary}.
\item
We next investigate the situation in case only linear rules are
allowed in the definition of the Datalog fragment used on the left
hand side of the containment problem (this distinction generally
makes no difference for the right-hand side). We find that in most
of these cases, the complexities mentioned above drop to
\kExpSpace{$(k+1)$}.
\end{itemize}

In summary, our results settle open problems for (nested) MQs, and
they paint a comprehensive and detailed picture of the state of the
art in Datalog query containment.

%
%

%
\section{Preliminaries}\label{sec:prelims}

We consider a standard language of first-order predicate logic, based on an infinite set \Ilang{} of \emph{constant symbols}, an infinite set \Plang{} of \emph{predicate symbols}, and an infinite set \Vlang{} of first-order \emph{variables}. Each predicate $p\in\Plang$ is associated with a natural number $\arity(p)$ called the \emph{arity} of $p$. The list of predicates and constants forms the language's \emph{signature} $\signature=\tuple{\Plang,\Ilang}$. We generally assume $\signature=\tuple{\Plang,\Ilang}$ to be fixed, and only refer to it explicitly if needed.

\paragraph*{Formulae, Rules, and Queries}
A \emph{term} is a variable $x\in\Vlang$ or a constant $c\in\Ilang$. We use symbols $s,t$ to denote terms, $x,y,z,v,w$ to denote variables, $a,b,c$ to denote constants. 
Expressions like $\vec{t}$, $\vec{x}$, $\vec{c}$ denote finite lists of such entities.
We use the standard predicate logic definitions of \emph{atom} and \emph{formula}, using symbols $\varphi$, $\psi$ for the latter.


Datalog queries are defined over an extended signature with additional predicate symbols, called \emph{IDB predicates}; all other predicates are called \emph{EDB predicates}.
A \emph{Datalog rule} is a formula of the form $\forall\vec{x},\vec{y}.\varphi[\vec{x},\vec{y}] \to\psi[\vec{x}]$ where $\varphi$ and $\psi$ are conjunctions of atoms, called the \emph{body} and \emph{head} of the rule, respectively, and where $\psi$ only contains IDB predicates. We usually omit universal quantifiers when writing rules. Sets of Datalog rules will be denoted by symbols $\mathbb{P},\mathbb{R},\mathbb{S}$. A set of Datalog rules $\mathbb{P}$ is
\begin{itemize}
\item \emph{monadic} if all IDB predicates are of arity one;
\item \emph{frontier-guarded} if the body of every rule
contains an atom $p(\vec{t})$ such that $p$ is an EDB predicate and $\vec{t}$ contains all variables
that occur in the rule's head;
\item \emph{linear} if every rule contains at most one IDB predicate in its body.
\end{itemize}

A \emph{conjunctive query} (\cq) is a formula $Q[\vec{x}]= \exists\vec{y}.\psi[\vec{x},\vec{y}]$ where $\psi[\vec{x},\vec{y}]$ is a conjunction of atoms;
a \emph{union of conjunctive queries} (\ucq) is a disjunction of such formulae.
A \emph{Datalog query} $\tuple{\mathbb{P},Q}$ consists of a set of Datalog rules $\mathbb{P}$ and a conjunctive query $Q$ over IDB or EDB predicates ($Q$ could be expressed as a rule in Datalog, but not in all restrictions of Datalog we consider). We write \datalog for the language of Datalog queries. A monadic Datalog query is one where $\mathbb{P}$ is monadic, and similarly for other restrictions. We use the query languages
\mdatalogconst (monadic), \gdatalog (frontier-guarded), \lindatalog (linear), and \linmdatalogconst (linear, monadic).


\paragraph*{Databases and Semantics}

We use the standard semantics of first-order logic (FOL).
A \emph{database instance} $\Inter$ consists of a set $\Delta^\Inter$ called \emph{domain} and a function $\cdot^\Inter$ that maps constants $c$ to domain elements $c^\Inter\in\Delta^\Inter$ and predicate symbols $p$ to relations $p^\Inter\subseteq (\Delta^\Inter)^{\arity(p)}$, where $p^\Inter$ is the \emph{extension} of $p$. 

Given a database instance $\Inter$ and a formula $\varphi[\vec{x}]$ with free variables $\vec{x}=\tuple{x_1,\ldots,x_m}$, the \emph{extension} of $\varphi[\vec{x}]$ is the subset of $(\Delta^\Inter)^m$ containing all those tuples $\tuple{\delta_1,\ldots,\delta_m}$ for which $\Inter,\{x_i \mapsto \delta_i \mid 1 \leq i \leq m\} \models \varphi[\vec{x}]$.
We denote this by $\tuple{\delta_1,\ldots,\delta_m}\in\varphi^\Inter$ or by $\Inter\models\varphi(\delta_1,\ldots,\delta_m)$; a similar notation is used
for all other types of query languages.
Two formulae $\varphi[\vec{x}]$ and $\psi[\vec{x}]$ are called \emph{equivalent} if their extensions coincide for every database instance $\Inter$.

The set of answers of a \ucq $Q[\vec{x}]$ over $\Inter$ is its extension.
The set of answers of a Datalog query $\tuple{\mathbb{P},Q}$ over $\Inter$ is the
intersection of the extensions of $Q$ over all extended database instances $\Inter'$ that
interpret IDB predicates in such a way that all rules of $\mathbb{P}$ are satisfied.
 Datalog \cite{Alice} can also be defined 
 as the least fixpoint on the inflationary evaluation of $Q$ on $I$.

Note that we do not require database instances to have a finite domain, since all of our results are valid in either case.
This is due to the fact that every entailment of a Datalog program has a finite witness,
and that all of our query languages are positive, i.e., that their answers are preserved under homomorphisms of
database instances.



%
\section{Guarded Queries}\label{sec:queries}

Monadically defined queries have been introduced in \cite{RK13:flagcheck} as a
generalization of monadic Datalog (\mdatalogconst) and conjunctive two-way regular path queries (\ctworpq{}s)
for which query containment is still decidable.\footnote{The queries were called $\qlang{MODEQ}$ in \cite{RK13:flagcheck}; we shorten this to \mq.}
The underlying idea of this approach is that candidate query answers are checked
by evaluating a monadic Datalog program, i.e., in contrast to the usual evaluation of
Datalog queries, we start with a ``guessed'' answer that is the input to a Datalog program.
To implement this, the candidate answer is represented by special constants $\flagconst$
that the Datalog program can refer to. This mechanism was called \emph{flag~\& check},
since the special constants act as flags to indicate the answer that should be checked.

\begin{example}\label{ex_transitivity}
A query that computes the transitive closure over a relation $p$ can be defined as follows.
%
\begin{align*}
p(\flagconst_1,y) & \to\colorpred(y)\\
\colorpred(y)\wedge p(y,z) & \to\colorpred(z)\\
\colorpred(\flagconst_2) & \to \checkpred
\end{align*}
One defines the answer of the query to contain all pairs $\tuple{\delta_1,\delta_2}$ for which the rules entail $\checkpred$ when interpreting $\flagconst_1$ as $\delta_1$ and $\flagconst_2$ as $\delta_2$.
\end{example}%

The approach used monadic Datalog for its close relationship to monadic second-order logic, which was the
basis for showing decidability of query containment. In this work, however, we develop new techniques for
showing the decidability (and exact complexity) of this problem directly. It is therefore suggestive
to consider other types of Datalog programs to implement the ``check'' part. The following definition
therefore introduces the general technique for arbitrary Datalog programs, and defines interesting
fragments by imposing further restrictions.

\begin{definition}\label{def_fcq}
Consider a signature $\signature$. An FCP (``flag \& check program'') of arity $m$ is a set of Datalog rules $\mathbb{P}$ with $k\geq 0$ IDB predicates $\colorpred_1,\ldots,\colorpred_k$, that may use the
additional constant symbols $\flagconst_1,\ldots,\flagconst_m\notin\signature$ and an additional nullary predicate symbol $\checkpred$.
An FCQ  (``flag \& check query'') $P$ is of the form $\exists\vec{y}.\mathbb{P}(\vec{z})$, where $\mathbb{P}$ is an FCP of arity $|\vec{z}|$ and all variables in $\vec{y}$ occur in $\vec{z}$.
The variables $\vec{x}$ that occur in $\vec{z}$ but not in $\vec{y}$ are the \emph{free variables} of $P$.

%

Let $\Inter$ be a database instance over $\signature$. The \emph{extension}
$\mathbb{P}^\Inter$ of $\mathbb{P}$ is the set
of all tuples $\tuple{\delta_1,\ldots,\delta_m}\in(\Delta^\Inter)^m$
such that every database instance $\Inter'$ that extends $\Inter$ to the signature of $\mathbb{P}$
and that satisfies $\tuple{\flagconst_1^{\Inter'},\ldots,\flagconst_m^{\Inter'}}=\tuple{\delta_1,\ldots,\delta_m}$ also entails $\checkpred$.
The semantics of FCQs is defined in the obvious way based on the extension of FCPs.
%

A \gq is an FCQ $\exists\vec{y}.\mathbb{P}(\vec{z})$ such that $\mathbb{P}$ is frontier-guarded.
Similarly, we define \mq (monadic), \linmq (linear, monadic), and \lingq (linear, frontier-guarded) queries.
\end{definition}

In contrast to \cite{RK13:flagcheck}, we do not define monadic queries as conjunctive queries of FCPs,
but we merely allow existential quantification to project some of the FCP variables.
Proposition~\ref{prop_posfcqs} below shows that this does not reduce expressiveness.

We generally consider monadic Datalog as a special case of frontier-guarded Datalog.
Monadic Datalog rules do not have to be frontier-guarded. A direct way to obtain
a suitable guard is to assume that there is a unary $\textsf{domain}$ predicate that
contains all (relevant) elements of the domain of the database instance. However, it
already suffices to require \emph{safety} of Datalog rules, i.e., that the variable in
the head of a rule must also occur in the body.
Then every element that is inferred to belong to an IDB relation must also occur in
some EDB relation. We can therefore add single EDB guard atoms to each rule in all
possible ways without modifying the semantics. This is a polynomial operation, since
all variables in the guards are fresh, other than the single head variable that we want to
guard. We therefore find, in particular, the \gq captures the expressiveness of \mq.
The converse is not true, as the following example illustrates.


\begin{example}\label{ex_gqvsmq}
The following $4$-ary \lingq generalizes Example~\ref{ex_transitivity} by checking for
the existence of two parallel $p$-chains of arbitrary length, where each
pair of elements along the chains is connected by a relation $q$, like the steps of
a ladder.
\begin{align*}
q(\flagconst_1,\flagconst_2) & \to\colorpred_q(\flagconst_1,\flagconst_2)\\
\colorpred_q(x,y)\wedge p(x,x')\wedge p(y,y'), q(x',y') & \to\colorpred_q(x',y')\\
\colorpred_q(\flagconst_3,\flagconst_4) & \to \checkpred
\end{align*}
One might assume that the following \mq is equivalent:
\begin{align*}
q(\flagconst_1,\flagconst_2) & \to\colorpred_1(\flagconst_1)\\
q(\flagconst_1,\flagconst_2) & \to\colorpred_2(\flagconst_2)\\
\colorpred_1(x)\wedge \colorpred_2(y)\wedge p(x,x')\wedge p(y,y'), q(x',y') & \to\colorpred_1(x')\\
\colorpred_1(x)\wedge \colorpred_2(y)\wedge p(x,x')\wedge p(y,y'), q(x',y') & \to\colorpred_2(y')\\
\colorpred_1(\flagconst_3)\wedge\colorpred_2(\flagconst_4) & \to \checkpred
\end{align*}
However, the latter query also matches structures that are not ladders.
For example, the following database yields the answer $\tuple{a,b,c,d}$,
although there is no corresponding ladder structure:
$\{q(a,b),p(a,c),p(b,e),q(c,e),p(a,e'),p(b,d),q(e',d)\}$.
One can extend the \mq to avoid this case, but any such fix is ``local''
in the sense that a sufficiently large ladder-like structure can trick
the query.
\end{example}%

It has been shown that monadically defined queries can be expressed both in
Datalog and in monadic second-order logic \cite{RK13:flagcheck}. While we lose
the connection to monadic second-order logic with \gq{}s, the expressibility in
Datalog remains. The encoding is based on the intuition that the choice of the
candidate answers for $\vec{\flagconst}$ ``contextualizes'' the inferences
of the Datalog program. To express this without special constants, we can store this
context information in predicates of suitably increased arity.

\begin{example}\label{ex_gqvsmqdatalog}
The $4$-ary \lingq of Example~\ref{ex_gqvsmq} can be expressed with the following
Datalog query. For brevity, let $\vec{y}$ be the variable list $\tuple{y_1,y_2,y_3,y_4}$,
which provides the context for the IDB facts we derive.
\begin{align*}
q(y_1,y_2) & \to\colorpred^+_q(y_1,y_2,\vec{y})\\
\colorpred_q(x,y,\vec{y})\wedge p(x,x')\wedge p(y,y'), q(x',y') & \to\colorpred^+_q(x',y',\vec{y})\\
\colorpred_q(y_3,y_4,\vec{y}) & \to \mathsf{goal}(\vec{y})
\end{align*}
This result is obtained by a straightforward extension of the translation algorithm
for \mq{}s \cite{RK13:flagcheck}, which may not produce the most concise representation.
Also note that the first rule in this program is not safe, since $y_3$ and $y_4$ occur
in the head but not in the body. According to the semantics we defined, such variables can be
bound to any element in the active domain of the given database instance (i.e., they
behave as if bound by a unary $\mathsf{domain}$ predicate).
\end{example}%

This observation justifies that we consider \mq{}s, \gq{}s, etc.\ as Datalog fragments.
It is worth noting that the translation does not change the number of IDB predicates
in the body of rules, and thus preserves linearity. The relation to (linear) Datalog also yields
some complexity results for query answering; we will discuss these
at the end of the next section, after introducing nested variants
our query languages.

\section{Nested Queries}\label{sec_nesting}

Every query language gives rise to a nested language, where we allow
nested queries to be used as if they were predicates. Sometimes, this does
not lead to a new query language (like for \cq and \datalog), but often it
affects complexities and/or expressiveness. It has been shown that both are
increased when moving from \mq{}s to their nested variants \cite{RK13:flagcheck}.
We will see that nesting also has strong effects on the complexity of query containment.

\begin{definition}
We define $k$-nested FCPs inductively. A $1$-nested FCP is an FCP. A $k+1$-nested FCP is an FCP that may use $k$-nested FCPs of arity $m$ instead of predicate symbols of arity $m$
in rule bodies. The semantics of nested FCPs is immediate based on the extension of FCPs.
A $k$-nested FCQ $P$ is of the form $\exists\vec{y}.\mathbb{P}(\vec{z})$, where $\mathbb{P}$ is a $k$-nested FCP of arity $|\vec{z}|$ and all variables in $\vec{y}$ occur in $\vec{z}$.

A $k$-nested \gq query is a $k$-nested frontier-guarded FCQ.
For the definition of \emph{frontier-guarded}, we still require EDB predicates in guards: subqueries cannot be guards. The language of $k$-nested \gq queries is denoted \kgq{k}; the language of arbitrarily nested \gq queries is denoted \ngq. Similarly, we define languages \kmq{k} and \nmq (monadic),
\klinmq{k} and \nlinmq (linear, monadic), and \klingq{k} and \nlingq (linear, frontier-guarded).
\end{definition}

Note that nested queries can use the same additional symbols (predicates and constants); this does not lead to
any semantic interactions, however, as the interpretation of the special symbols is ``private'' to each query.
To simplify notation, we assume that distinct (sub)queries always contain distinct special symbols.
The relationships of the query languages we introduced here are summarized in Figure~\ref{fig_querycompl},
where upwards links denote increased expressiveness. An interesting observation that is represented in
this figure is that linear Datalog is closed under nesting:

\begin{theorem}\label{theo_lindatalognesting}
\lindatalog = \nestedq{\linq{\datalog}}.
\end{theorem}

%

Another kind of nesting that does not add expressiveness is the nesting of FCQs in \ucq{}s.
Indeed, it turns out that (nested) FCQs can internalize arbitrary conjunctions and disjunctions
of FCQs (of the same nesting level). This even holds when restricting to linear rules.

\begin{proposition}\label{prop_posfcqs}
Let $P$ be a positive query, i.e., a Boolean expression of disjunctions and conjunctions, of
\klinmq{k} queries with $k\geq 1$.
Then there is a \klinmq{k} query $P'$ of size polynomial in $P$ that is equivalent to $P$.
Analogous results hold when replacing \klinmq{k} by \kmq{k}, \kgq{k}, or \klinmq{k} queries.
\end{proposition}

Query answering for \mq{}s has been shown to be
\NP-complete (combined complexity) and \PTime-complete (data complexity).
For \nmq, the combined complexity increases to \PSpace while the data
complexity remains the same. These results can be extended to frontier-guarded queries.
We also note the query complexity for frontier-guarded Datalog, for which we are
not aware of any published result.

\begin{theorem}\label{theo_gqquerycomlpexity}
The combined complexity of evaluating \gq queries over a database
instance is \NP-complete. The same holds for \gdatalog queries.
The combined complexity of evaluating \ngq queries is \PSpace-complete.
The data complexity is \PTime-complete for \gdatalog, \gq, and \ngq.
\end{theorem}

The lower bounds in the previous case are immediate from know results
for monadically defined queries. In particular, the hardness proof for nested \mq{}s
also shows that queries of a particular fixed nesting level can encode the
validity problem for quantified boolean formulae with a certain number
of quantifier alternations; this explains why we show the combined complexity of \kmq{k}
to be in the Polynomial Hierarchy in Figure~\ref{fig_querycompl}.
A modification of this hardness proof from \cite{RK13:flagcheck} allows us to
obtain the same results for the combined complexities in the linear cases;
matching upper bounds follow from Theorem~\ref{theo_gqquerycomlpexity}.

\begin{theorem}\label{theo_linquerycomlpexity}
The combined complexity of evaluating \linmq queries over a database
instance is \NP-complete. The same holds for \lingdatalog{} and \lingq.
The combined complexity of evaluating \nlinmq queries is \PSpace-complete.
The same holds for \nlingq.

The data complexity is \NLogSpace-complete for all of these query languages.
\end{theorem}

%

\section{Deciding Query Containment with Automata}\label{sec_containmentautomata}

We first recall a general technique of reducing query containment
to the containment problem for (tree) automata \cite{ChaudhuriV97},
which we build our proofs on. An introduction to tree automata is included
in the appendix.

A common way to describe the answers of a \datalog query $P=\tuple{\aprogram,p}$ is to consider its
\emph{expansion trees}. Intuitively speaking, the goal atom $p(\vec{x})$ can be rewritten by applying
rules of $\aprogram$ in a backward-chaining manner until all IDB predicates have been eliminated,
resulting in a \cq. The answers of $P$ coincide with the (infinite) union of answers
to the \cq{}s obtained in this fashion. The rewriting itself gives rise to a tree structure,
where each node is labeled by the instance of the rule that was used in the rewriting, and
the leaves are instances of rules that contain only EDB predicates in their body.
The set of all expansion trees provides a regular description of $P$ that
we exploit to decide containment.

To formalize this approach, we describe the set of all expansion trees as a tree language,
i.e., as a set of trees with node labels from a finite alphabet. The number of possible labels
of nodes in expansion trees is unbounded, since rules are instantiated using fresh variables.
To obtain a finite alphabet of labels, one limits the number of variables and thus the overall
number of possible rule instantiations \cite{ChaudhuriV97}.

\begin{definition}\label{defn_prooftree}
Given a \datalog query $P=\tuple{\aprogram,p}$,
$\mathcal{R}_{\aprogram}$ is the set of all instantiations of rules of $\aprogram$
using only the variables $\mathcal{V}_{\aprogram}=\{v_1,\ldots,v_n\}$, where $n$ is twice the maximal number of variables
occurring in any rule of $\aprogram$.

A \emph{proof tree} for $P$ is a tree with labels from $\mathcal{R}_{\aprogram}$, such
that
(a) the root is labeled by a rule with $p$ as its head predicate;
(b) if a node is labeled by a rule $\arule$ with an IDB atom $B$ in its body,
then it has a child node that is labeled by $\arule'$ with head atom $B$.
The label of a node $e$ is denoted $\pi(e)$.

Consider two nodes $e_1$ and $e_2$ in a proof tree with lowest common ancestor $e$.
Two occurrences of a variable $v$ in $\pi(e_1)$ and $\pi(e_2)$ are \emph{connected}
if $v$ occurs in the head of $\pi(f)$ for all nodes $f$ on the shortest path
between $e_1$ and $e_2$, with the possible exception of $e$.
\end{definition}

A proof tree encodes an expansion tree where we replace every set of mutually
connected variable occurrences by a fresh variable.
Conversely, every expansion tree
is represented by a proof tree that replaces fresh body variables by variables that
do not occur in the head; this is always possible since proof trees can use twice
as many variables as any rule of $\aprogram$.
The set of proof trees is a regular tree language that can be described by an automaton.

\begin{proposition}[Proposition 5.9 \cite{ChaudhuriV97}]\label{prop_taprooftree}
For a \datalog query $P=\tuple{\aprogram,p}$, there is a
tree automaton $\aautomaton_P$ of size exponential in $P$
that accepts exactly the set of all proof trees of $P$.
\end{proposition}

In order to use $\aautomaton_P$ to decide containment of $P$ in
another query $P'$, we construct an automaton $\aautomaton_{P\sqsubseteq P'}$
that accepts all proof trees of $P$ that are ``matched'' by $P'$.
Indeed, every proof tree induces a \emph{witness}, i.e., a minimal matching database
instance, and one can check whether or not $P'$ can produce the same query answer on this instance.
If this is the case for all proof trees of $P$, then containment is shown.

\section{Deciding Guarded Query Containment}\label{sec_gqcontainmentup}

Our first result provides the upper bound for deciding containment
of \gq queries. In fact, the result extends to arbitrary \datalog queries
on the left-hand side.

\begin{theorem}\label{theo_membdlgq}
\containmentMembershipStatement{\datalog}{\gq}{\ThreeExpTime}
\end{theorem}

To prove this, we need to construct the tree automaton $\aautomaton_{P\sqsubseteq P'}$
for an arbitrary \gq $P'$.
As a first step, we construct
an alternating 2-way tree automaton $\aautomaton^+_{P\sqsubseteq P'}$ that accepts the
proof trees that we would like $\aautomaton_{P\sqsubseteq P'}$ to accept, but
with nodes additionally being annotated with information about the choice of $\flagconst$ values
to guide the verification.

We first construct automata to verify the match of a single, non-recursive rule that may
refer to $\flagconst$ constants. The rule does not have to be monadic or frontier-guarded.
Our construction is inspired by a similar construction
for \cq{}s by Chaudhuri and Vardi \cite{ChaudhuriV97}, with the main difference that
the answer variables in our case are not taken from the root of the tree but rather from
one arbitrary node that is marked accordingly.

To define this formally, we introduce trees with additional annotations besides their node labels. Clearly,
such trees can be viewed as regular labelled trees by considering annotations to be components of one label;
our approach, however, leads to a more readable presentation.

\begin{definition}\label{defn_annotree}
Consider a Datalog program $\aprogram$, a rule $\arule=\varphi\to p(\vec{x})$,
and $n\geq 0$ special constants $\vec{\flagconst}=\flagconst_1,\ldots,\flagconst_n$.
The proof-tree variables $\mathcal{V}_{\aprogram}$ used in $\mathcal{R}_{\aprogram}$
are as in Definition~\ref{defn_prooftree}.

A proof tree for $\aprogram$ is \emph{$\vec{\flagconst}$-annotated} if
every node has an additional \emph{$\vec{\flagconst}$-label} that is a partial mapping $\{\flagconst_1,\ldots,\flagconst_n\}\to\mathcal{V}_{\aprogram}$,
such that:
every special constant $\flagconst_i$ occurs in at least one $\vec{\flagconst}$-label,
and whenever a constant $\flagconst_i$ occurs in two $\vec{\flagconst}$-labels, it is
mapped to the same variable and both variable occurrences are connected.

A proof tree for $\aprogram$ is \emph{$p$-annotated} if exactly one node
has an additional \emph{$p$-label} of the form $p(\vec{v})$, where $\vec{v}$ is a list of variables
from $\mathcal{V}_{\aprogram}$.

A \emph{matching tree} $T$ for $\arule$ and $\aprogram$ is a $\vec{\flagconst}$-annotated
and $p$-annotated proof tree for $\aprogram$ for which there is a mapping $\nu:\Var(\arule)\cup\{\flagconst_1,\ldots,\flagconst_n\}\to\mathcal{V}_{\aprogram}$ such that
\begin{enumerate}[(a)]
\item $\nu(p(\vec{x}))=p(\vec{v})$;
\item for every atom $\alpha$ of $\varphi$, there is a node $e_\alpha$ in $T$ such that the rule instance that $e_\alpha$ is labeled with contains the EDB atom $\nu(\alpha)$ in its body;
\item if $\flagconst_i$ occurs in $\alpha$, then the $\vec{\flagconst}$-label maps $\flagconst_i$ to the occurrence of $\nu(\flagconst_i)$ in $e_\alpha$;
\item if $\alpha,\alpha'\in\varphi$ share a variable $x$, then the occurrences of $\nu(x)$ in $e_\alpha$ and $e_{\alpha'}$ are connected.
\end{enumerate}
\end{definition}

\begin{proposition}\label{prop_ruleauto}
There is an automaton $\aautomaton_{P,\arule}$ that accepts exactly the annotated matching trees for $\arule$ and $\aprogram$,
and which is exponential in the size of $\arule$ and $\aprogram$.
\end{proposition}

We want to use the automata $\aautomaton_{P,\arule}$ to verify the entailment of a single rule within a
Datalog derivation. We would like an automaton to check whether a whole derivation is possible.
Unfortunately, we cannot check these derivations using automata of
the form $\aautomaton_{P,\arule}$, which each need to be run on a $p$-annotated tree which has the unique
entailment of the rule marked. The length of a derivation is unbounded, and we would not be able to
distinguish an unbounded amount of $p$-markers.
To overcome this problem, we create a modified automaton $\aautomaton^+_{P,\arule,\vec{v}}$ that simulates the
behavior of $\aautomaton_{P,\arule}$ on a tree with annotation $p(\vec{v})$. For $\aautomaton^+_{P,\arule,\vec{v}}$ to
know which node the annotation $p(\vec{v})$ refers to, it has to be started at this node.
This is a non-standard notion of run, where we do not start at the root of the tree.
Moreover, starting in the middle of the tree makes it necessary to consider both nodes below and above
the current position, and $\aautomaton^+_{P,\arule,\vec{v}}$ therefore needs to be an \emph{alternating 2-way tree automaton}.

\begin{proposition}\label{prop_ruleautoplus}
There is an alternating 2-way tree automaton $\aautomaton^+_{P,\arule,\vec{v}}$ that is polynomial in the size of
$\aautomaton_{P,\arule}$ such that, whenever $\aautomaton_{P,\arule}$ accepts a matching tree $T$ that has the $p$-annotation
$p(\vec{v})$ on node $e$, then $\aautomaton^+_{P,\arule,\vec{v}}$ has an accepting run that starts from the corresponding
node $e'$ on the tree $T'$ that is obtained by removing the $p$-annotation from $T$.
\end{proposition}

Using the automata $\aautomaton^+_{P,\arule,\vec{v}}$, we can now obtain the claimed alternating 2-way automaton
$\aautomaton^+_{P\sqsubseteq P'}$ for a \gq $P'$. Intuitively speaking, $\aautomaton^+_{P\sqsubseteq P'}$
concatenates the automata $\aautomaton^+_{P,\arule,\vec{v}}$ using alternation: whenever a derivation requires
a (recursive) IDB atom, a suitable process $\aautomaton^+_{P,\arule,\vec{v}}$ is initiated, starting from a node
in the middle of the tree. The construction relies on guardedness, which ensures that we can always
find a suitable start node (corresponding to the node that was $p$-annotated earlier), by finding a suitable
guard EDB atom in the tree.

\begin{proposition}\label{prop_containmentaltauto}
For a \datalog query $P$ and a \gq query $P'$ with special constants $\vec{\flagconst}$, there is an alternating 2-way automaton
$\aautomaton^+_{P\sqsubseteq P'}$ of exponential size that accepts
the $\vec{\flagconst}$-annotated proof trees of $P$ that encode expansion trees with $\vec{\flagconst}$ assignments for
which $P'$ has a match.
\end{proposition}

We are now ready to prove Theorem~\ref{theo_membdlgq}.
The automaton $\aautomaton^+_{P\sqsubseteq P'}$ allows us to check the answers of
$P'$ on a proof tree that is $\vec{\flagconst}$-annotated to assign values for answer constants.
We can transform this alternating 2-way automaton into a tree automaton
$\aautomaton'_{P\sqsubseteq P'}$ that is exponentially larger, i.e., doubly exponential in the size of
the input.
To remove the need for $\vec{\flagconst}$-labels, we modify the automaton $\aautomaton'_{P\sqsubseteq P'}$
so that it can only perform a transition from its start state if it finds that the constants
in $\vec{\flagconst}$ are assigned to the answer variables of $P$ in the root.
Finally, we obtain $\aautomaton_{P\sqsubseteq P'}$ by projecting to the alphabet $\mathcal{R}_{\aprogram}$
without $\vec{\flagconst}$-annotations; this is again possible in polynomial effort.
The containment problem $P\sqsubseteq P'$ is equivalent by deciding the containment of
$\aautomaton_P$ in $\aautomaton_{P\sqsubseteq P'}$, which is possible in exponential time w.r.t.\ to the
size of the automata. Since $\aautomaton_P$ is exponential and $\aautomaton_{P\sqsubseteq P'}$ is
double exponential, we obtain the claimed triple exponential bound.

Our proof of Theorem~\ref{theo_membdlgq} can be used to obtain another interesting result
for the case of frontier-guarded Datalog. If $P$ is a \gdatalog query, which does not
use any special constants $\flagconst$, then the $\vec{\flagconst}$-annotations are not
relevant and $\aautomaton^+_{P\sqsubseteq P'}$ can be constructed as an alternating 2-way automaton
on proof trees.
For this, we merely need to modify the construction in Proposition~\ref{prop_containmentaltauto}
to start in start states of automata for rules that entail the goal predicate of $P'$ with the expected binding of
variables to answer variables of $P$.
We can then omit the projection step, which required us to convert
$\aautomaton^+_{P\sqsubseteq P'}$ into a tree automaton earlier.
Instead, we can construct from $\aautomaton^+_{P\sqsubseteq P'}$ a complement tree
automaton $\bar{\aautomaton}_{P\sqsubseteq P'}$ that is only exponentially larger than $\aautomaton^+_{P\sqsubseteq P'}$, i.e., doubly exponential overall \cite{Cosmadakis88}[Theorem~A.1]. Containment can
then be checked by checking the non-emptiness of $\aautomaton_P\cap \bar{\aautomaton}_{P\sqsubseteq P'}$,
which is possible in polynomial time, leading to a \TwoExpTime{} algorithm overall.

\begin{theorem}\label{theo_membdlgdl}
\containmentMembershipStatement{\datalog}{\gdatalog}{\TwoExpTime}
\end{theorem}

This generalizes an earlier result of Cosmadakis et al.\ for
monadic Datalog \cite{Cosmadakis88} using an alternative, direct proof.


Finally, we can lift our results to the case of nested queries.
Using Proposition~\ref{prop_posfcqs}, we can make the simplifying assumption
that rules with some nested query in their body contain only one
nested query and a guard atom as the only other atom. Thus all rules with
nested queries have the form $g(\vec{s})\wedge Q(\vec{t})\to p(\vec{u})$, where
$g$ is an EDB predicate, $Q$ is a nested query, and the variables $\vec{u}$ occur in $\vec{s}$.

In Proposition~\ref{prop_ruleautoplus}, we constructed alternating 2-way automata
$\aautomaton^+_{P,\arule,\vec{v}}$ that can check the entailment of a particular
atom $p(\vec{v})$ starting from a node within the tree. Analogously,
we now construct automata $\aautomaton^+_{P,Q,\theta}$ that check that the nested query
$Q$ matches partially, where $\theta$ is a substitution that interprets
query variables in terms of proof-tree variables on the current node of the tree.
Only the variables that occur in $g(\vec{s})$ and $Q(\vec{t})$ are mapped by $\theta$; the
remaining variables can be interpreted arbitrarily, possibly in distant parts of the proof tree.

To construct $\aautomaton^+_{P,Q,\theta}$, we use the alternating 2-way automaton
$\aautomaton^+_{P\sqsubseteq Q}$, constructed in Proposition~\ref{prop_containmentaltauto}
(assuming, for a start, that $Q$ is not nested). This automaton is extended to an alternating 2-way automaton
$\aautomaton^+_{P,Q}$ that accepts trees with a unique annotation of the form
$\tuple{Q,\theta}$, for which we check that it is consistent with the $\vec{\flagconst}$-annotation
(i.e., for each query variable $x$ mapped by $\theta$, the corresponding constant $\flagconst$
is assigned to $\theta(x)$ at the node that is annotated with $\tuple{Q,\theta}$).
We then obtain a (top-down) tree automaton $\aautomaton_{P,Q}$ by transforming
$\aautomaton^+_{P,Q}$ into a tree automaton (exponential), and projecting away the
$\vec{\flagconst}$-annotations (polynomial). The automaton $\aautomaton_{P,Q}$ is analogous to
the tree automaton $\aautomaton_{P,\arule}$ of Proposition~\ref{prop_ruleauto}.
Using the same transformation as in Proposition~\ref{prop_ruleautoplus}, we obtain
an alternating 2-way automaton $\aautomaton^+_{P,Q,\theta}$ for each $\theta$.

The automaton $\aautomaton^+_{P\sqsubseteq P'}$ for a nested query $P'$ is constructed as in
Proposition~\ref{prop_containmentaltauto}, but using the automata $\aautomaton^+_{P,Q,\theta}$
instead of automata $\aautomaton^+_{P,\arule,\vec{v}}$ to check the entailment of
a subquery $Q$. The size of $\aautomaton^+_{P\sqsubseteq P'}$ is increased by one exponential
since the size of $\aautomaton^+_{P,Q,\theta}$ is exponentially increased when projecting out
$\vec{\flagconst}$-labels for $Q$. Applying this construction inductively, we obtain the
following result.

\begin{theorem}\label{theo_membdlgqk}
\containmentMembershipStatement{\datalog}{\kgq{k}}{\kExpTime{$(k+2)$}}
\end{theorem}

%

\section{Simulating Alternating Turing Machines}\label{sec_atmencoding}

\begin{figure*}
\graphicspath{{figures/}}
\mbox{}\hfill\scalebox{0.85}{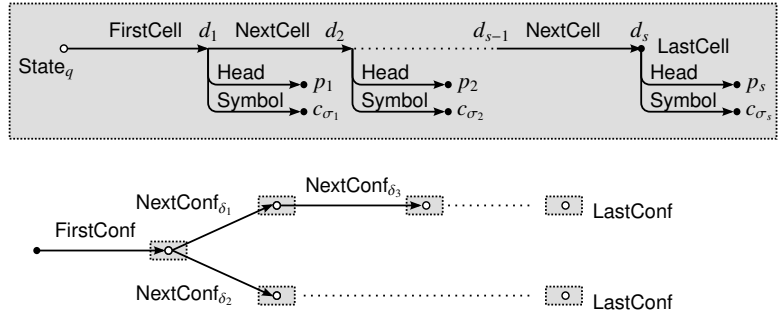}\hfill\mbox{}
\caption{Illustration of the ATM encoding of Definition~\ref{def_atmencode}: shaded configurations (top) are used within the configuration tree (bottom);
$\textsf{ConfCell}$ queries are omitted for clarity\label{fig_atmencode}}
\end{figure*}

To show the hardness of query containment problems, we generally provide direct
encodings of Alternating Turing Machines (ATMs) with a fixed space bound \cite{ATM}.
To simplify this encoding, we assume without loss of generality that
every universal ATM configuration leads to exactly two successor configurations.
The following definition defines ATM encodings formally. Rather than requiring
concrete structures to encode ATMs, we abstract the encoding by means of queries
that find suitable structures in a database instance; this allows us to apply
the same definition for increasingly complex encodings.
The following definition is illustrated in Figure~\ref{fig_atmencode}.

\begin{definition}\label{def_atmencode}
Consider an ATM $\mathcal{M}=\tuple{Q,\Sigma,\Delta,q_s,q_e}$ and queries
$\textsf{FirstConf}[x,y]$, $\textsf{NextConf}_\delta[x,y]$ for all $\delta\in\Delta$, $\textsf{LastConf}[x]$,
$\textsf{State}_q[x]$ for all $q\in Q$,
$\textsf{Head}[x,y]$,
$\textsf{ConfCell}[x,y]$, $\textsf{FirstCell}[x,y]$, $\textsf{NextCell}[x,y]$, $\textsf{LastCell}[x]$,
and $\textsf{Symbol}[x,y]$. To refer to tape symbols, we consider constants $c_\sigma$ for all $\sigma\in\Sigma$,
and to refer to positions of the head, we use constants $h$ (here), $l$ (left), and $r$ (right).

With respect to these queries, an element $c\in\domain{\Inter}$ in a database instance $\Inter$
\emph{encodes an $\mathcal{M}$ quasi-configuration of size $s$} if $\Inter$ contains a structure\medskip

\noindent{\small%
\[%
\begin{array}{@{}l}
\textsf{State}_q(c), \textsf{FirstCell}(c,d_1),
\\\textsf{ConfCell}(c,d_1), \textsf{Symbol}(d_1,c_{\sigma_1}), \textsf{Head}(d_1,p_1), \textsf{NextCell}(d_1,d_2),\\
\textsf{ConfCell}(c,d_2), \textsf{Symbol}(d_2,c_{\sigma_2}), \textsf{Head}(d_2,p_2),\ldots, \textsf{NextCell}(d_{s-1},d_s),\\
\textsf{ConfCell}(c,d_s), \textsf{Symbol}(d_s,c_{\sigma_s}), \textsf{Head}(d_s,p_s),\textsf{LastCell}(d_s),
\end{array}
\]}%
where $q\in Q$, $\sigma_i\in\Sigma$, and $p_i\in\{h,l,r\}$.
We say that $c$ \emph{encodes an $\mathcal{M}$ configuration of size $s$} if, in addition, the sequence $(p_i)_{i=1}^s$ has the form
$l,\ldots,l,h,r,\ldots,r$
with zero or more occurrences of $r$ and $l$, respectively.

An element $c$ in $\Inter$ \emph{encodes a (quasi-)configuration tree of $\mathcal{M}$ in space $s$} if
\begin{itemize}
\item $\Inter\models\textsf{FirstConf}(c,d_1)$ for some $d_1$,
\item $d_1$ is the root of a tree with edges defined by $\textsf{NextConf}_\delta$,
\item every node in this tree encodes an $\mathcal{M}$ \mbox{(quasi-)}\allowbreak{}configuration of size $s$,
\item if there is a transition $\Inter\models\textsf{NextConf}_{\delta_1}(e,e_1)$, where $\delta_1=\tuple{q,\sigma,q',\sigma',d}$
and $q$ is a universal state, then there is also a transition $\Inter\models\textsf{NextConf}_{\delta_2}(e,e_2)$ with $\delta_1\neq\delta_2$,
\item if $e$ is a leaf node, then $\Inter\models\textsf{LastConf}(e)$.
\end{itemize}
If the tree is an accepting run, then $c$ encodes an accepting run (of $\mathcal{M}$ in space $s$).


A \emph{same-cell query} is a query $\textsf{SameCell}[x,y]$ such that, if $c_1,c_2\in\domain{\Inter}$
encode two quasi-configurations, and $d_1,d_2\in\domain{\Inter}$ represent the same tape cell
in the encodings $c_1$ and $c_2$, respectively, then $\tuple{d_1,d_2}\in\textsf{SameCell}^\Inter$.


Two queries $P_1[x]$ and $P_2[x]$ \emph{containment-encode} accepting runs of $\mathcal{M}$ in space $s$ if, for every
database instance $\Inter$ and element $c\in P_1^\Inter\setminus P_2^\Inter$,
$c$ encodes an accepting run of $\mathcal{M}$ in space $s$, and every accepting run of
$\mathcal{M}$ in space $s$ is encoded by some $c\in P_1^\Inter\setminus P_2^\Inter$ for some $\Inter$.
\end{definition}

Note that elements $c$ may encode more than one configuration (or configuration tree). This is
not a problem in our arguments.

The conditions that ensure that a quasi-configuration tree is an accepting run can be expressed
by a query, based on the queries given in Definition~\ref{def_atmencode}. More specifically,
one can construct a query that accepts all elements that encode a quasi-configuration sequence that is \emph{not} a run.
Together with a query that accepts only encodings of quasi-configurations tree,
this allows us to containment-encode accepting runs of an ATM.
Only linear queries, possibly nested, will be needed to perform the required checks, even in the case of ATMs.
To simplify the statements, we use \klinmq{0} as a synonym for \ucq.

\begin{lemma}\label{lemma_atmquasienctoenc}
Consider an ATM $\mathcal{M}$, and queries as in Definition~\ref{def_atmencode}, including $\textsf{SameCell}[x,y]$,
that are \kmq{k} queries for some $k\geq 0$.
There is a \kmq{k} query $P[x]$, polynomial in the size of $\mathcal{M}$ and the given queries, such that the following hold.
\begin{itemize}
\item For every accepting run of $\mathcal{M}$ in space $s$, there is some database instance $\Inter$ with some element $c$ that encodes the run, such that $c\notin P^\Inter$.
\item If an element $c$ of $\Inter$ encodes a tree of quasi-configurations of $\mathcal{M}$ in space $s$, and if $c\notin P^\Inter$,
then $c$ encodes an accepting run of $\mathcal{M}$ in space $s$.
\end{itemize}
Moreover, if all input queries are in \klinmq{k}, then so is $P$.
\end{lemma}

The previous result allows us to focus on the encoding of quasi-configuration trees and the definition of queries
as required in Definition~\ref{def_atmencode}. Indeed, the main challenge below will be to enforce a sufficiently
large tape for which we can still find a correct same-cell query.

\section{Hardness of Monadic Query Containment}\label{sec_mqcontainmentlow}

We can now prove our first major hardness result:

\begin{theorem}\label{theo_hardmdlmqk}
\containmentHardnessStatement{\mdatalogconst}{\kmq{k}}{\kExpTime{$(k+2)$}}
\end{theorem}

Note that the statement includes the \ThreeExpTime-hardness for containment of \mq{}s as a special case.
%
%
To prove this result, we first construct an \ExpSpace ATM that we then use to
construct tapes of double exponential size.

\begin{lemma}\label{lemma_mducqexpatm}
For any ATM $\mathcal{M}$, there is an \mdatalogconst query $P_1[x]$, a \linmq $P_2[x]$, queries as in Definition~\ref{def_atmencode}
that are \linmq{}s, and a same-cell query that is a \ucq, such that $P_1[x]$ and $P_2[x]$ containment-encode accepting
runs of $\mathcal{M}$ in exponential space.
\end{lemma}

\begin{figure*}[t!]
\graphicspath{{figures/}}
\mbox{}\hfill\scalebox{0.85}{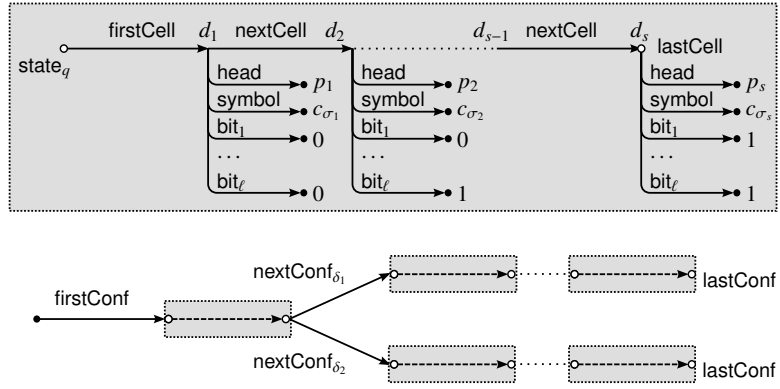}\hfill\mbox{}
\caption{Illustration of the ATM encoding of Lemma~\ref{lemma_mducqexpatm}: shaded configurations (top) are used within the configuration tree (bottom)\label{fig_mducqexpatm}}
\end{figure*}

Figure~\ref{fig_mducqexpatm} illustrates the encoding that we use to prove Lemma~\ref{lemma_mducqexpatm}.
While it resembles the structure of Figure~\ref{fig_atmencode}, the labels are now EDB predicates rather than
(abstract) queries. The encoding of tapes attaches to each cell an $\ell$-bit address (where bits are represented
by constants $0$ and $1$). We can use these bits to count from $0$ to $2^\ell$ to construct tapes of
this length. The query on the left-hand side can only enforce that there are cells with bit addresses, not
that they actually count; even the exact length of the tape is unspecified. The query on the right-hand side
of the containment then checks that consecutive cells (in all tapes that occur in
the configuration tree) represent successor addresses, and that the first and last address is as expected.

Another difference from Figure~\ref{fig_atmencode} is that we now treat configurations as linear structures, with a
beginning and an end. In our representation of the configuration tree, we next configuration therefore connects to the
last cell of the previous configuration's tape, rather than its start. We do this to ensure that the encoding
works well even when restricting to linear queries. Indeed, the only non-linear rules in $P_1$ are used to enforce
multiple successor configurations for universal states of an ATM. For normal TMs, even $P_1$ is in \linmdatalogconst{}.
The rules of the $P_1$ are as follows:\smallskip

\noindent{\small%
\begin{align*}
\textsf{firstConf}(x,y)\wedge\colorpred_{\textit{conf}}(y) &\to\colorpred_{\textit{goal}}(x)\\
%
\textsf{state}_q(x)\wedge\textsf{firstCell}(x,y)\wedge\colorpred_{\textit{bit}_1}(y) &\to\colorpred_{\textit{conf}}(x) & \text{for $q\in Q$}\displaybreak[0]\\
\textsf{bit}_{i-1}(x,0)\wedge\colorpred_{\textit{bit}_i}(x) &\to\colorpred_{\textit{bit}_{i-1}}(x) & \text{for $i\in\{2,\ldots,\ell\}$}\displaybreak[0]\\
\textsf{bit}_{i-1}(x,1)\wedge\colorpred_{\textit{bit}_i}(x) &\to\colorpred_{\textit{bit}_{i-1}}(x) & \text{for $i\in\{2,\ldots,\ell\}$}\displaybreak[0]\\
\textsf{symbol}(x,c_\sigma)\wedge\colorpred_{\textit{symbol}}(x) &\to\colorpred_{\textit{bit}_\ell}(x) & \text{for $\sigma\in\Sigma$}\displaybreak[0]\\
\textsf{head}(x,p)\wedge\colorpred_{\textit{head}}(x) &\to\colorpred_{\textit{symbol}}(x) &\text{for $p\in\{h,r,l\}$}\displaybreak[0]\\
%
%
\textsf{nextCell}(x,y)\wedge\colorpred_{\textit{bit}_1}(y) &\to\colorpred_{\textit{head}}(x) & \displaybreak[0]\\
\textsf{nextConf}_\delta(x,y)\wedge\colorpred_{\textit{conf}}(y) &\to\colorpred_{\textit{head}}(x) & \text{for $\delta=\tuple{q,\sigma,q',\sigma',d}$}\\[-0.7ex]
 && \text{with $q\in Q_\exists$}\displaybreak[0]\\
%
\textsf{nextConf}_{\delta_1}(x,y_1)\wedge\colorpred_{\textit{conf}}(y_1) \wedge{}&  &\text{for $\delta_1=\tuple{q,\sigma,q',\sigma',d}$, }\\[-0.7ex]
\textsf{nextConf}_{\delta_2}(x,y_2)\wedge\colorpred_{\textit{conf}}(y_2) &\to\colorpred_{\textit{head}}(x) & \text{$q\in Q_\forall$, and $\delta_1\neq\delta_2$} \displaybreak[0]\\
\textsf{lastConf}(x) &\to\colorpred_{\textit{head}}(x)
\end{align*}%
}%
Note that we do not enforce any structure to define the query $\textsf{ConfCell}$;
this query is implemented by a \linmq{} that navigates over an arbitrary number of cells
within one configuration. This is the main reason why we need \linmq{}s rather than
\ucq{}s here.

We now use the exponential space ATM of Lemma~\ref{lemma_mducqexpatm} to encode the tape
of \kExpSpace{2} ATM.
The following result shows, that one can always obtain an exponentially larger tape
by nesting linear queries on the right-hand side.

\begin{lemma}\label{lemma_atmencodenesting}
Assume that there is some space bound $s$ such that, for every DTM $\mathcal{M}$, there is a
\mdatalogconst query $P_1[x]$ and an \kmq{k+1} query $P_2[x]$ with $k\geq 0$, such that
$P_1[x]$ and $P_2[x]$ containment-encode accepting runs of $\mathcal{M}$ in $s$,
where the queries required by Definition~\ref{def_atmencode} are \kmq{k+1} queries.
Moreover, assume that there is a suitable same-cell query that is in \kmq{k}.

Then, for every ATM $\mathcal{M}'$, there is a \mdatalogconst query $P_1'[x]$, an \kmq{k+1} $P_2'[x]$,
and \kmq{k+1} queries as in Definition~\ref{def_atmencode}, such that
$P_1'[x]$ and $P_2'[x]$ containment-encode an accepting run of $\mathcal{M}'$ in space $s'\geq 2^s$.
Moreover, the size of the queries for this encoding is polynomial in the size of the queries for the original encoding.
\end{lemma}

We show this result by using a deterministic space-$s$ Turing machine $\mathcal{M}$ to count from $0$ to $2^s$,
which takes a fixed number $s'>2^s$ of steps. We then use the encodings of accepting runs of
$\mathcal{M}$ as encodings for tapes of the ATM $\mathcal{M}'$, where every configuration of $\mathcal{M}$
becomes a cell of $\mathcal{M}'$. All tapes simulated in this way are of equal length $s'$.
Some queries required by Definition~\ref{def_atmencode} are easy to obtain: for example, the
new query $\textsf{NextCell}'[x,y]$ is the query $\textsf{NextConf}[x,y]$ of the encoding of $\mathcal{M}$.
The most difficult to express is the new same-cell query, for which we use the following \kmq{k+1}:

\noindent{\small%
\begin{align*}
\textsf{FirstCell}(\lambda_1,x) &\to\colorpred_1(x)\\
\colorpred_1(x)\wedge\textsf{NextCell}(x,x') &\to\colorpred_1(x')\\
%
\textsf{State}_q(\lambda_1)\wedge\textsf{FirstCell}(\lambda_1,x)\wedge\textsf{Symbol}(x,z)\wedge\textsf{Head}(x,v)\wedge{}\\[-0.7ex]
\textsf{State}_q(\lambda_2)\wedge\textsf{FirstCell}(\lambda_2,y)\wedge\textsf{Symbol}(y,z)\wedge\textsf{Head}(y,v) &\to\colorpred_2(y)\\[-0.7ex]
	& \phantom{{}\to{}}\text{for all $q\in Q$}\\
\colorpred_1(x)\wedge\colorpred_2(y)\wedge\textsf{SameCell}(x,y)\wedge{}\\[-0.7ex]
\textsf{NextCell}(x,x')\wedge\textsf{Symbol}(x',z)\wedge\textsf{Head}(x',v)\wedge{}\\[-0.7ex]
\textsf{NextCell}(y,y')\wedge\textsf{Symbol}(y',z)\wedge\textsf{Head}(y',v) &\to\colorpred_2(y')\\
\colorpred_2(y)\wedge \textsf{LastCell}(y)&\to\checkpred
\end{align*}}%
where $\textsf{FirstCell}$, $\textsf{Symbol}$, $\textsf{SameCell}$, and $\textsf{LastCell}$ are the queries
from the encoding of $\mathcal{M}$.
The first two rules simply mark the tape starting at $\flagconst_1$ with $\colorpred_1$.
The next two rules then compare the two (potentially very long) tapes from configurations of $\mathcal{M}$ to check
if they contain exactly the same symbols at each position, and the last rule finishes.
Since the tapes are not connected in any known way,
we have to be careful to ensure never to loose the connection to
either of the tapes, to avoid comparing random cells from other parts of the database. Indeed,
the last two rules do not mention $\flagconst_1$ or $\flagconst_2$ at all.
We need two IDB predicates to achieve this, which carefully mark the two tapes cell by cell.

Another important thing to note is that the query $\textsf{SameCell}$ is only used
exactly once in exactly one rule. Indeed, if we were using it twice, then the length
of our queries would grow exponentially when applying the construction inductively.
This is the reason why we encode symbols and head positions with constants, rather than
using unary predicates like for states. In the latter case, we need many rules, one for
each predicate, as can be seen in the third rule above.
One could try to avoid the use of constants by more complex encodings that
encode information using paths of different lengths as done by
Bj\"{o}rklund et al.~\cite{BMS08:xmlschemacont}. However, some additional device is needed
to ensure that database instances are sufficiently closely connected in this case, which
may again require constants, IDBs of higher arity, or a greater nesting level of \linmq queries
to navigate larger distances.

With the previous results, Theorem~\ref{theo_hardmdlmqk} can be proved by an easy induction:
for the base case $k=1$ we apply Lemma~\ref{lemma_atmencodenesting}
to the result of Lemma~\ref{lemma_mducqexpatm}; for the induction step
we use Lemma~\ref{lemma_atmencodenesting} again.

\section{Linear Datalog}\label{sec:linearity}

%

Not only query answering, but also containment checking is often slightly
simpler in fragments of linear Datalog. Intuitively, this is so because
derivations can be represented as words rather than as trees. Thus, the
automata theoretic techniques that we have used in Section~\ref{sec_gqcontainmentup}
can be applied with automata on words where some operations are easier.
In particular, containment of (nondeterministic) automata on words can be
checked in polynomial space rather than in exponential time.
This allows us to establish the following theorems, which reduce
the \kExpTime{$2$} upper bound of Theorem~\ref{theo_membdlgdl} to \ExpSpace and
the \kExpTime{$(k+2)$} upper bound of Theorem~\ref{theo_membdlgqk} to \kExpSpace{$(k+1)$}.


\begin{theorem}\label{theo_memblindlgdl}
\containmentMembershipStatement{\lindatalog}{\gdatalog}{\ExpSpace}
\end{theorem}

\begin{theorem}\label{theo_memblindlklingq}
\containmentMembershipStatement{\lindatalog}{\kgq{k}}{\kExpSpace{$(k+1)$}}
\end{theorem}


Establishing matching lower bounds for the complexity turns out to be more difficult.
In general, we loose the power of alternation, which explains the reduction in complexity.
The general approach of encoding (non-alternating) Turing machines is the same as in Section~\ref{sec_atmencoding},
where Definition~\ref{def_atmencode} is slightly simplified since we do not need to
consider universal states, so that configuration trees turn into configuration sequences.
Moreover, Lemma~\ref{lemma_atmquasienctoenc} applies to this case as well, since it only requires
linear queries. Likewise, our general inductive step in Lemma~\ref{lemma_atmencodenesting}
uses deterministic (non-alternating) TMs to construct exponentially long tapes.
Moreover, it turns out that the construction of an initial exponential space TM in  Lemma~\ref{lemma_mducqexpatm}
leads to linear queries if the TM has no universal states.

Yet it is challenging to lift the exact encodings of Lemma~\ref{lemma_mducqexpatm} and Lemma~\ref{lemma_atmencodenesting}.
The same-cell query that we constructed in Lemma~\ref{lemma_atmencodenesting} for our inductive argument
is non-linear. As explained in Section~\ref{sec_mqcontainmentlow}, the use of two IDBs
to mark both sequences of tape cells is essential there to ensure correctness.
The main problem is that we must not loose connection to either of the sequences during our
checks.
As an alternative to using IDBs on both sequences, one could use the $\textsf{ConfCell}$ query to ensure that the
compared cells belong to the right configurations. This leads to the following same-cell query:

\noindent{\small%
\begin{align*}
%
\textsf{State}_q(\lambda_1)\wedge\textsf{FirstCell}(\lambda_1,x)\wedge\textsf{Symbol}(x,z)\wedge\textsf{Head}(x,v)\wedge{}\\[-0.7ex]
\textsf{State}_q(\lambda_2)\wedge\textsf{FirstCell}(\lambda_2,y)\wedge\textsf{Symbol}(y,z)\wedge\textsf{Head}(y,v) &\to\colorpred(y)\\[-0.7ex]
	& \phantom{{}\to{}}\text{for all $q\in Q$}\\
\colorpred(y)\wedge\textsf{ConfCell}(\lambda_1,x)\wedge\textsf{SameCell}(x,y)\wedge{}\\[-0.7ex]
\textsf{NextCell}(x,x')\wedge\textsf{Symbol}(x',z)\wedge\textsf{Head}(x',v)\wedge{}\\[-0.7ex]
\textsf{NextCell}(y,y')\wedge\textsf{Symbol}(y',z)\wedge\textsf{Head}(y',v) &\to\colorpred(y')\\
\colorpred(y)\wedge \textsf{LastCell}(y)&\to\checkpred
\end{align*}}
While this works in principle, it has the problem that the $\textsf{ConfCell}$ query of Lemma~\ref{lemma_mducqexpatm}
is a \linmq, not a \ucq. Therefore, if we construct a same-cell query for the \kExpSpace{$2$} case,
we obtain \klinmq{2} queries, which yields the following result:

\begin{theorem}\label{theo_hardlinmqklinmq}
\containmentHardnessStatement{\linmdatalogconst}{\klinmq{k}}{\kExpSpace{$k$}}
\end{theorem}

In order to do better, one can try to express $\textsf{ConfCell}$ as a \ucq.
In general, this is not possible on the database instances that the left-hand query
in Lemma~\ref{lemma_mducqexpatm} recognizes, since cells may have an exponential distance
to their configuration while \ucq{}s can only recognize local structures.
To make $\textsf{ConfCell}$ local, we can modify the left-hand query to ensure that every
cell is linked directly to its configuration with a binary predicate $\textsf{inConf}$.
Using binary IDB predicates, we can do this with the following set of frontier-guarded rules:

\noindent{\small%
\begin{align*}
\textsf{firstConf}(x,y)\wedge\colorpred_{\textit{conf}}(y) &\to\colorpred_{\textit{goal}}(x)\\
\textsf{state}_q(x)\wedge\textsf{nextCell}(x,y)\wedge{}&\\[-0.7ex]
\textsf{inConf}(y,x)\wedge\colorpred_{\textit{bit}_1}(y,x) &\to\colorpred_{\textit{conf}}(x) & \text{for $q\in Q$}\displaybreak[0]\\
\textsf{bit}_{i-1}(x,0)\wedge\colorpred_{\textit{bit}_i}(y,z)\wedge\textsf{inConf}(x,z) &\to\colorpred_{\textit{bit}_{i-1}}(x,z) & \text{for $i\in\{2,\ldots,\ell\}$}\displaybreak[0]\\
\textsf{bit}_{i-1}(x,1)\wedge\colorpred_{\textit{bit}_i}(y,z)\wedge\textsf{inConf}(x,z) &\to\colorpred_{\textit{bit}_{i-1}}(x,z) & \text{for $i\in\{2,\ldots,\ell\}$}\displaybreak[0]\\
\textsf{symbol}(x,c_\sigma)\wedge\colorpred_{\textit{symbol}}(x,z)\wedge\textsf{inConf}(x,z) &\to\colorpred_{\textit{bit}_\ell}(x,z) & \text{for $\sigma\in\Sigma$}\displaybreak[0]\\
\textsf{head}(x,h)\wedge\colorpred_{\textit{head}}(x,z)\wedge\textsf{inConf}(x,z) &\to\colorpred_{\textit{symbol}}(x,z) &\displaybreak[0]\\
\textsf{head}(x,l)\wedge\colorpred_{\textit{head}}(x,z)\wedge\textsf{inConf}(x,z) &\to\colorpred_{\textit{symbol}}(x,z) &\displaybreak[0]\\
\textsf{head}(x,r)\wedge\colorpred_{\textit{head}}(x,z)\wedge\textsf{inConf}(x,z) &\to\colorpred_{\textit{symbol}}(x,z) & \displaybreak[0]\\
\textsf{nextCell}(x,y)\wedge\colorpred_{\textit{bit}_1}(y,z)\wedge\textsf{inConf}(x,z) &\to\colorpred_{\textit{head}}(x,z) & \displaybreak[0]\\
\textsf{nextConf}_\delta(x,y)\wedge\colorpred_{\textit{conf}}(y)\wedge\textsf{inConf}(x,z) &\to\colorpred_{\textit{head}}(x,z) & \text{for $\delta\in\Delta$} \displaybreak[0]\\
\textsf{lastConf}(x)\wedge\textsf{inConf}(x,z) &\to\colorpred_{\textit{head}}(x,z) \\
\end{align*}%
}%
Structures matched by this query provide direct links from each element to their configuration
element, and we can thus formulate $\textsf{ConfCell}$ as a \ucq and obtain the following.

\begin{theorem}\label{theo_hardlingqklinmq}
\containmentHardnessStatement{\lingdatalog}{\klinmq{k}}{\kExpSpace{$(k+1)$}}
\end{theorem}

It is not clear if this result can be extended to containments of \linmq in \klinmq{k};
the above approach does not suggest any suitable modification. In particular, the
propagation of $\textsf{inConf}$ in the style of a transitive closure does not work,
since elements may participate in many $\textsf{inConf}$ relations.
On the other hand, the special constants $\flagconst$ in \linmq{}s cannot be used to
refer to the current configuration, since there can be an unbounded number of configurations
but only a bounded number of special constants. It is possible, however, to
formulate a \linmq $\textsf{Config}[x]$ that generates the required structure for a single configuration,
since one can then represents the configuration by $\flagconst$. We can generate arbitrary
sequences of such structures by using $\textsf{Config}[x]$ as a nested query to
that matches a regular expression $\textsf{firstConf}~(\textsf{Config}~\textsf{NextConf})^\ast~\textsf{Config}~\textsf{lastConf}$,
where we use $\textsf{NextConf}$ to express the disjunction of all $\textsf{nextConf}_\delta$ relations.
This proves the following statement.

\begin{theorem}\label{theo_hardtwolinmqklinmq}
\containmentHardnessStatement{\klinmq{2}}{\klinmq{k}}{\kExpSpace{$(k+1)$}}
\end{theorem}

Finally, we can also continue to use the same approach for encoding $\textsf{SameCell}$
as in Section~\ref{sec_mqcontainmentlow}, without using $\textsf{ConfCell}$, while still
restricting to linear Datalog (and thus to non-alternating TMs) on the left-hand side.
This leads us to the following result.

\begin{theorem}\label{theo_hardlinmdlmq}
\containmentHardnessStatement{\linmdatalogconst}{\kmq{k}}{\kExpSpace{$(k+1)$}}
\end{theorem}

We have thus established tight complexity bounds for the containment of
nested \gq{}s, while there remains a gap (of one exponential or one nesting level)
for \mq{}s.

%
%

%
%
\section{Conclusions}\label{sec:conc}

\begin{table*}
\mbox{}\hfill
\scalebox{0.9}{\!\!\!\!\begin{tabular}{|l|c|c|c|c|c|} \hline
     & \ucq,               &             & &  & \\
  & \linmdatalogconst{},  \mdatalogconst,  & \klinmq{k},  &  \kmq{k}, & \nlinmq,\nmq, &  \\
  & \lingdatalog{},   \gdatalog  & \klingq{k}  &   \kgq{k} & \nlingq,\ngq & \datalog \\
\hline \linmq
 &   \PSpace-h \cite{ChaudhuriV94} & \kExpSpace{$k$}-h  [Th.\ref{theo_hardlinmqklinmq}] &  \kExpSpace{$(k+1)$}-c & Nonelementary & Undecidable \\
&     \ExpSpace [Th.\ref{theo_memblindlgdl}]       &    \kExpSpace{$(k+1)$} [Th.\ref{theo_memblindlklingq}]   &   [Th.\ref{theo_hardlinmdlmq}]$\setminus$[Th.\ref{theo_memblindlklingq}] & [Th.\ref{theo_hardlinmqklinmq}] & \cite{Alice} \\
\hline \lingdatalog,
 &  \ExpSpace-c & \kExpSpace{$(k+1)$}-c  & \kExpSpace{$(k+1)$}-c  & Nonelementary & Undecidable \\
\klinmq{n} ($n \geq 2$),
 &  &  &  &  & \\
\nlinmq, \nlingq,
 &     [Th.\ref{theo_hardtwolinmqklinmq}]$\setminus$[Th.\ref{theo_memblindlgdl}]       &       [Th.\ref{theo_hardlingqklinmq},\ref{theo_hardtwolinmqklinmq}]$\setminus$[Th.\ref{theo_memblindlklingq}] &
 [Th.\ref{theo_hardlingqklinmq},\ref{theo_hardtwolinmqklinmq}]$\setminus$[Th.\ref{theo_memblindlklingq}]  &  [Th.\ref{theo_hardlingqklinmq},\ref{theo_hardtwolinmqklinmq}] & \cite{Alice} \\

\klingq{n}, \lindatalog
 & & &  &  &  \\
\hline
\hline \mdatalogconst, \gdatalog,
                   &  \TwoExpTime-c                                  & \kExpTime{$(k+2)$}-c &                    \kExpTime{$(k+2)$}-c  & Nonelementary & Undecidable \\
 \kmq{n}$\!\!$, \kgq{n}$\!\!$, & \cite{BenediktBS12,ChaudhuriV97}$\setminus$       &  [Th.\ref{theo_hardmdlmqk}]$\setminus$[Th.\ref{theo_membdlgqk}]          &  [Th.\ref{theo_hardmdlmqk}]$\setminus$[Th.\ref{theo_membdlgqk}]    & [Th.\ref{theo_hardmdlmqk}]& \cite{Shm87}\\
 \nmq$\!\!$, \ngq$\!\!$, \datalog         & \cite{Cosmadakis88}, [Th.\ref{theo_membdlgdl}] &                                         &  & & \\
\hline
\end{tabular}

}
\hfill\mbox{}
\caption{Summary of the known complexities of query containment for several Datalog fragments; sources for each claim
are shown in square brackets, using $\setminus$ to separate sources for lower and upper complexity bounds, respectively \label{fig_contcompl}}
\end{table*}
We have studied the most expressive fragments of Datalog for which
query containment is still known to be decidable today, and we have provided exact
complexities for most of their query answering and query containment problems.
While containment for nested queries tends to be non-elementary for unbounded nesting depth, we have
shown tight exponential complexity hierarchies for the main cases that we studied.
As part of our results, we have also settled a number of open problems for
known query languages: the complexity of query containment for \mq and \nmq,
the complexity of query containment of \datalog in \gdatalog, and the expressivity of nested \lindatalog.

Moreover, we have built on the recent ``flag~\& check'' approach of monadically defined queries to
derive various natural extensions, which lead to new query languages with interesting
complexity results. In most cases, we observed that the extension from monadic to
frontier-guarded Datalog does not affect any of the complexities, whereas it might
have an impact on expressivity. In contrast, the restriction to linear Datalog has
the expected effects, both for query answering and for query containment.

The only case for which our results for containment complexity are not tight is when we restrict
rules to be both linear and monadic: while small variations in the involved query languages
lead to the expected tight bounds, this particular combination eludes our analysis.
This case could be studied as part of a future program for analyzing the behavior of (nested)
conjunctive regular path queries, which are also a special form of monadic, linear Datalog.

Another interesting open question is the role of constants. Our hardness proofs, especially
in the nested case, rely on the use of constants to perform certain checks more efficiently.
Without this, it is not clear how an exponential blow-up of our encoding (or the use of additional
nesting levels) could be avoided. Of course, constants can be simulated if we have either predicates of
higher arity or special constants as in ``flag \& check'' queries. However, for the case
of (linear) monadic Datalog without constants, we conjecture that containment complexities
are reduced by one exponential each when omitting constants.

An additional direction of future research is to study problems where we ask for the \emph{existence} of
a containing query of a certain type rather than merely check containment of two given queries.
The most prominent instance of this scenario is the \emph{boundedness} problem, which asks
whether a given Datalog program can be expressed by some (yet unknown) \ucq. It has been shown that
this problem can be studied using tree-automata-based techniques as for query containment \cite{Cosmadakis88},
though other approaches have been applied as well \cite{BaranyCO12}.
Besides boundedness, one can also ask more general questions of \emph{rewritability}, e.g., whether
some Datalog program can be expressed in monadic Datalog or in a regular path query.

%
%
\section{References}
{\small
\bibliographystyle{abbrv}
\newcommand\refname{}
\renewcommand{\section}[2][\empty]{}
\bibliography{references}
}


\clearpage
\appendix

\section{Tree Automata}

We use standard definitions for two-way alternating tree automata as introduced in \cite{Cosmadakis88}.
A regular (one-way, non-alternating) tree automaton is obtained by restricting
this definition.

Tree automata run over ranked, labelled trees of some maximal arity (out-degree) $f$.
A ranked tree can be seen a function $t$ mapping sequences of positive natural numbers
(encoding nodes in the tree) to symbols from a fixed finite alphabet (the labels of each node).
Each letter of the alphabet is ranked, i.e., associated with an arity that defines
how many child nodes a node labeled with this symbol should have.
The domain of $t$, denoted $\nodes(t)$, satisfies the following closure property:
if $\sq{i}\cdot j\in\nodes(t)$, then $\sq{i}\in\nodes(t)$
and $\sq{i}\cdot k\in\nodes(t)$ for all $1\le k\le j$.
Given a ranked tree $t$,
we write $\sq{i}\in\nodes(t)$ to denote an arbitrary node of $t$
and $t(\sq{i})$ to denote the label of $\sq{i}$ in $t$.
We denote by $\trees(\Sigma)$ the set of trees over the alphabet $\Sigma$.

A two-way alternating tree automaton $\aautomaton$ is a tuple
$\tuple{\Sigma,Q,Q_s,\delta,Q_e}$ where
\begin{itemize}
\item $\Sigma$ is a tree alphabet;
\item $Q$ is a set of states;
\item $Q_s\subseteq Q$ is the set of initial states;
\item $Q_e\subseteq Q$ is the set of accepting states;
\item $\delta$ is a transition function from $Q \times \Sigma$:
let $q\in Q$ be a state and $\sigma\in\Sigma$ be a letter of arity $\ell$;
then $\delta(q,\sigma)$ is a positive boolean combination of elements in
$ \{-1,0,1, \cdots,\ell\}\times Q$.
\end{itemize}

The numbers used in transitions encode directions, where $-1$ is up and $0$ is stay.
For example $\delta(q,\sigma)=(\tuple{1,s_1} \wedge \wedge{1,s_2}) \vee (\tuple{-1,t_3} \wedge \tuple{2,t_4})$ is an example
of transition for a state $q$ and a node labeled $\sigma$:
a node labeled by $\sigma$ can be in the state $q$ iff
its first child can be in the states $s_1$ and $s_2$,
or its parent and its second child can be in the states $s_3$ and $s_4$, respectively.

Let $t$ be a tree over $\Sigma$.
A run $\tau$ of $\aautomaton$ over $t$ is a tree labeled by elements of $Q \times \{-1,0,1,\cdots,f\} \times \nodes(t)\cup \{-1\}$.
$\tau$ satisfies the following properties:
\begin{itemize}
\item $\tau$ is finite.
\item The root of $\tau$ is labelled by $(q_0,i,n)$, where $q_0$ is in $Q_s$.
\item If a node $v$ is labelled by $(q,i,n)$ and $n$ is not a node of $t$,
then $v$ is a leaf of $\tau$.
\item If a node $v$ is labelled by $(q,i,n')$, $n$ is a node of $\tau$ labelled by $\sigma$ of arity $l$ and $v'$ is labelled by $(q_1,j,n')$ then
\begin{itemize}
\item if $j=-1$, then there exists $u \leq k$ such that $n = n'.u$
\item if $j=0$, then $n = n'$
\item if $j \leq k$, then $n'= n.j$.
\end{itemize}
\item if a node $v$ is labelled by $(q,i,n)$, $n \in t$ labelled by $\sigma$
and the children of $v$
are labelled by $(q_1,j_1,n_1) \cdots (q_k,j_k,n_k)$ then $\delta(q,\sigma)$
is satisfied when interpreting the sybmols $\{\tuple{j_1,q_1}, \cdots, \tuple{j_k,q_k}\}$
as \emph{true} and all other symbols as \emph{false}.
\end{itemize}

$\tau$ is \emph{valid} iff, for each leaf of $\tau$ labelled by $(q,i,n)$, $q$ is in $Q_e$.
$\aautomaton$ \emph{accepts} a tree $t$ if
there exists a valid run of $t$ over $\aautomaton$.
We denote by $\trees(\aautomaton)$. The set of trees accepted by $\aautomaton$.

A regular (one-way, non-alternating) tree automaton is a 2-way alternating tree automaton
where all transitions for a symbol $\sigma$ of rank $\ell$ are boolean formulae of the
form $(\tuple{1,q_{11}}\wedge\ldots\wedge\tuple{\ell,q_{\ell1}})\vee\ldots\vee(\tuple{1,q_{1n}}\wedge\ldots\wedge\tuple{\ell,q_{\ell{}n}})$
for some $n\geq 0$. In particular, directions $0$ and $-1$ do not occur. In this case, we can represent
transitions as sets of lists of states
$\{\tuple{q_{11},\ldots,q_{\ell1}},\ldots,\tuple{q_{1n},\ldots,q_{\ell{}n}}\}$.

Finally, we recall two useful theorems from \cite{Cosmadakis88}.

\begin{theorem}[Theorem A.1 of \cite{Cosmadakis88}]
Let $\aautomaton$ be a two-way alternating automaton.
Then there exists a tree automaton $\aautomaton’$ whose size is
exponential in the size of $\aautomaton$ such that
$\trees(\aautomaton’) = \trees(\Sigma)\setminus\trees(\aautomaton)$.
\end{theorem}

\begin{theorem}[Theorem A.2 of \cite{Cosmadakis88}]
Let $\aautomaton$ be a two-way alternating automaton.
Then there exists a tree automaton $\aautomaton’$ whose size is
exponential in the size of $\aautomaton$ such that
$\trees(\aautomaton’) = \trees(\aautomaton)$.
\end{theorem}

\section{Proofs}

\section*{Proofs for Section~\ref{sec_nesting}}

\begin{reptheorem}{theo_lindatalognesting}
\lindatalog = \nestedq{\linq{\datalog}}.
\end{reptheorem}
\begin{proof}
We will prove that any \nestedq{\linq{\datalog}} query can be
rewritten into a \lindatalog query of polynomial size. We make
simplifying assumptions on the structure of the nested query which
can be easily obtained by polynomial transformations and make the
presentation easier: we assume that every rule body of any query
occurring at any nesting depth contains at most one subquery atom
(using, e.g., Proposition~\ref{prop_posfcqs}).
Second, we assume that all variables and IDB predicates that are not
in the same scope are appropriately renamed apart.

In order to proof our claim, we will first show that any
\knestedq{\linq{\datalog}}{2} can be rewritten into an equivalent
\lindatalog query. Applying the rewriting iteratively inside-out
(and observing that even manyfold application can be done in
polynomial total time) then allows to conclude that there is a
polynomial rewriting of any \nestedq{\linq{\datalog}} query of
arbitrary depth into a \lindatalog query.

Consider a \knestedq{\linq{\datalog}}{2} query
$P=\tuple{\aprogram,p}$ and assume w.l.o.g. that every rule body of
the rules contains at most one \knestedq{\linq{\datalog}}{1}
subquery. Now, going through all rules of $\aprogram$ we produce the
rules $\aprogram'$ of the unnested but equivalent version.

Consider a rule $\rho\in \aprogram$ having the shape
$$
Q(x_1,\ldots,x_n) \wedge p(y_1,\ldots y_\ell) \wedge B_1 \wedge
\ldots \wedge B_k \to H
$$
where $p$ is the body IDB predicate and where
$Q=\tuple{\mathbb{Q},q}$ is a \knestedq{\linq{\datalog}}{1} query.
For any $k$-ary IDB predicate $r$ inside $\mathbb{Q}$ we increase
its arity by $\ell$ and let $\aprogram'$ contain all rules of
$\mathbb{Q}'$ which is obtained from the rules $\rho'$ of
$\mathbb{Q}$ by
\begin{itemize}
\item replacing any (head or body) IDB atom $r(z_1,\ldots,z_k)$ of $\rho'$
by $r(z_1,\ldots,z_k,y_1,\ldots y_\ell)$ and
\item in case $\rho'$ does not contain any IDB body atom, add $p(y_1,\ldots
y_\ell)$ to the body.
\end{itemize}
Further we let $\aprogram'$ contain the rule
$$
q(x_1,\ldots,x_n,y_1,\ldots y_\ell) \wedge \wedge B_1 \wedge \ldots
\wedge B_k \to H.
$$

In case of a rule $\rho\in \aprogram$ having the shape
$$
Q(x_1,\ldots,x_n) \wedge B_1 \wedge \ldots \wedge B_k \to H
$$
we add $\mathbb{Q}$ to $\aprogram'$ without change and let
$\aprogram'$ contain the rule
$$
q(x_1,\ldots,x_n) \wedge B_1 \wedge \ldots \wedge B_k \to H.
$$

In case a rule $\rho\in \aprogram$ does not contain a subquery atom
we simply add $\rho$ to $\aprogram'$.

It can now easily verified that $\tuple{\aprogram,p}$ and
$\tuple{\aprogram',p}$ are equivalent: first it is straightforward,
that $\tuple{\aprogram,p}$ is equivalent to
$\tuple{\aprogram^\flat,p}$ where $\aprogram^\flat$ is obtained from
$\aprogram$ by replacing every $Q(x_1,\ldots,x_n)$ by
$q(x_1,\ldots,x_n)$ (that is, the according goal predicate) and then
adding all rules from $\mathbb{Q}$ with no changes made to them.
Second one can show that there is a direct correspondence between
proof trees of $\tuple{\aprogram^\flat,p}$ and linearized proof
trees of $\tuple{\aprogram',p}$ which yields the desired result.
\end{proof}

\begin{repproposition}{prop_posfcqs}
Let $P$ be a positive query, i.e., a Boolean expression of disjunctions and conjunctions, of
\klinmq{k} queries with $k\geq 1$.
Then there is a \klinmq{k} query $P'$ of size polynomial in $P$ that is equivalent to $P$.
Analogous results hold when replacing \klinmq{k} by \kmq{k}, \kgq{k}, or \klinmq{k} queries.
\end{repproposition}
\begin{proof}
We show the claim by induction, by expressing the innermost disjunctions and conjunctions of $P$
with equivalent \klinmq{k} queries of linear size. We consider positive queries without existential
quantifiers (i.e., where all variables are answer variables), but the inner \klinmq{k} may use
existential quantifiers.

Let $P[\vec{x}]=P_1[\vec{x_1}]\vee \ldots\vee P_n[\vec{x_n}]$ be a disjunction of \klinmq{k} queries.
Each query $P_i$ is of the form $\exists\vec{z_i}.P'_i[\vec{x'_i}]$, where $\vec{x'_i}$
is the list of free variables of $P'_i$ (corresponding to constants $\flagconst$), and $\vec{z_i}$ contains exactly
those variables of $\vec{x'_i}$ that do not occur in $\vec{x_i}$. We assume without loss of generality
that $\vec{z_i}$ is disjoint from $\vec{z_j}$ if $i\neq j$, and that each $P'_i$ uses a unique set of IDBs
that does not occur in other queries.
We consider queries $\bar{P}_i$ obtained by replacing the special constant that represents a variable
$x_j\in\vec{x}$ by the special constant $\flagconst_j$ (assumed to not occur in $P$ yet). Thus,
the queries $\bar{P}_i$ share special constants exactly where queries $P_1$ share variables.
We can now define the \klinmq{k} $P'$ as $\exists\vec{z_1}\ldots\vec{z_n}.\bar{P}_1\cup\ldots\cup \bar{P}_n$,
where we assume that the correspondence of special constants to free variables is such that the
existential quantifiers refer to the same variables as before.

Let $P[\vec{x}]=P_1[\vec{x_1}]\wedge \ldots\wedge P_n[\vec{x_n}]$ be a conjunction of \klinmq{k} queries.
Let $P_i=\exists\vec{z_i}.P'_i[\vec{x'_i}]$ as before, and let $\colorpred_i$ for $i\in\{1,\ldots,n-1\}$ be
fresh IDB predicates.
The queries $\bar{P}_i$ are defined as before by renaming special constants to reflect shared variables.
For each $i\in\{1,\ldots,n\}$, the set of rules $\hat{P}_i$ is obtained from $\bar{P}_i$ as follows:
if $i<n$, then every rule $\varphi\to\checkpred\in \bar{P}_i$ is replaced by the rule $\varphi\to\colorpred_i(\flagconst_1)$, where $\flagconst_1$ is a fixed special constant in the queries;
if $i>1$, then every rule $\varphi\to \psi\in\bar{P}_i$ where $\varphi$ does not contain an IDB predicate
is replaced by the rule $\varphi\wedge\colorpred_{i-1}(\flagconst_1)\to \psi$, where $\flagconst_1$ is as before. The \klinmq{k} $P'$ is defined as $\exists\vec{z_1}\ldots\vec{z_n}.\hat{P}_1\cup\ldots\cup \hat{P}_n$.

These constructions lead to equivalent \klinmq{k} queries of linear size, so the claim follows by inductions.
The cases for \kmq{k}, \kgq{k}, and \klinmq{k} follow from the same constructions (note that, without the requirement of linearity, a simpler construction is possible in the case of conjunctions).
\end{proof}

\begin{reptheorem}{theo_gqquerycomlpexity}
The combined complexity of evaluating \gq queries over a database
instance is \NP-complete. The same holds for \gdatalog queries.
The combined complexity of evaluating \ngq queries is \PSpace-complete.
The data complexity is \PTime-complete for \gdatalog, \gq, and \ngq.
\end{reptheorem}
\begin{proof}
The lower bounds are immediate from the
matching complexities for \mq and \nmq queries, respectively \cite{RK13:flagcheck}.

First, we prove that checking if a tuple is an answer
of a \gq over a database instance $\Inter$ is in \NP
for combined complexity.
Let $\Inter$ be an instance, let $P$ be a \gq with frontier guarded
rules $\mathbb{P}$, and let $\vec{\delta}$ be
be a candidate answer for $P$ as in Definition~\ref{def_fcq}.

Since each rule in $\mathbb{P}$ is frontier-guarded, each
intentional fact that is derived when checking the answer follows
from the application of one particular rule, instantiated to
match one particular (guard) EDB fact in the body. Therefore,
the number of IDB facts that can be derived is polynomially bounded
in the size of $\Inter$ and $\mathbb{P}$.

Thus, for every derivation of $\mathbb{P}$, only a polynomial number
of rule applications are necessary, since it is enough to derive each
IDB fact once. It is clear that one can guess such a derivation,
where we guess, for each derivable IDB fact, one specific rule instance
by which it is derived. The correctness of this guess can be checked in
polynomial time, showing that the problem can be solved in \NP.
%

We now show that checking an answer of a \ngq over an instance $\Inter$ is in \PSpace.
Let $\Inter$ be an instance, let $P$ be a \kgq{k} with frontier guarded
rules $\mathbb{P}$ (that may contain subqueries), and let $\vec{\delta}$ be
be a candidate answer for $P$ as in Definition~\ref{def_fcq}.
We demonstrate by induction on $k$ that checking if
$\vec{\delta}$ is a solution for $P$ w.r.t.\ $\Inter$ is in \NPSpace.
For the induction base, the claim follows from the above result for \gq{}s.

For the induction step, using the same argument as before, we can see that
the number of IDB facts that can be derived by $\mathbb{P}$ is still polynomial.
Therefore, we can again guess a polynomial derivation as before, though the
rule instances now may refer to subqueries of smaller nesting depth.
By the induction hypothesis, whenever we need to verify the applicability of
such a rule, we can use an \NPSpace algorithm for the nested query.
The overall number of such checks is polynomial, yielding the overall \NPSpace algorithm.
The result follows since \NPSpace=\PSpace \cite{Savitch1970}.


The fact that query evaluation is in \PTime for data complexity is immediate from the
fact our queries can be expressed in Datalog, which is known to have this data complexity.
A direct proof is also obtained by observing that the number of possible derivation
sequences that the above algorithms need to consider is in itself polynomial in $\Inter$
if $P$ is fixed, so that the algorithms themselves are already in \PTime for data complexity.
%
\end{proof}

\begin{reptheorem}{theo_linquerycomlpexity}
The combined complexity of evaluating \linmq queries over a database
instance is \NP-complete. The same holds for \lingdatalog{} and \lingq.
The combined complexity of evaluating \nlinmq queries is \PSpace-complete.
The same holds for \nlingq.

The data complexity is \NLogSpace-complete for all of these query languages.
\end{reptheorem}
\begin{proof}
The claimed \NP-completeness is immediate. Hardness follows from the hardness
of \cq query answering. Membership follows from the membership of \gq.

The claimed membership in \PSpace follows from the \PSpace-membership of \lindatalog;
note that this uses Theorem~\ref{theo_lindatalognesting}.
Hardness for \nlingq follows from the hardness for \nlinmq, which we show by modifying
the \PSpace-hardness proof for monadically defined queries from \cite{RK13:flagcheck}.

\newcommand{\predstyle}[1]{\mathit{#1}}
\newcommand{\predabbr}{FCP}
We show the result by providing a reduction from the validity
problem of quantified Boolean formulae (QBFs). We recap that for any
QBF, it is possible to construct in polynomial time an equivalent
QBF that has the specific shape $$Q_1 x_1 Q_2 x_2 \ldots Q_n x_n
\bigvee_{L\in\mathcal{L}} \bigwedge_{\ell\in L} \ell,$$ with
$Q_1,\ldots Q_n \in \{\exists,\forall\}$ and $\mathcal{L}$ being a
set of sets of literals over the propositional variables
$x_1,\ldots,x_n$. In words, we assume our QBF to be in prenex form
with the propositional part of the formula in disjunctive normal
form. For every literal set $L = \{x_{k_1},\ldots,x_{k_i},$ $\neg
x_{k_{i+1}},\ldots, \neg x_{k_j}\}$, we now define the $n$-ary
\predabbr{} $\predstyle{p}_L = \{
\predstyle{t}(\flagconst_{k_1})\wedge\ldots\wedge\predstyle{t}(\flagconst_{k_i})\wedge
\predstyle{f}(\flagconst_{k_{i+1}})\wedge\ldots\wedge\predstyle{f}(\flagconst_{k_j})
\to \checkpred\}$. Moreover, we define the $n$-ary \predabbr{}
$\predstyle{p}_\mathcal{L} = \{
\predstyle{p}_L(\flagconst_1,\ldots,\flagconst_n)\to \checkpred \mid
L\in\mathcal{L} \}$. Letting $\predstyle{p}_\mathcal{L} =
\predstyle{p}_n$ we now define \predabbr{}s $\predstyle{p}_{n-1}
\ldots \predstyle{p}_{0}$ in descending order. If $Q_i = \exists$,
then the $i{-}1$-ary \predabbr{} $\predstyle{p}_{i-1}$ is defined as
the singleton rule set $\{
\predstyle{p}_{i}(\flagconst_1,\ldots,\flagconst_{i-1},y) \to
\checkpred \}$. In case $Q_i = \forall$, we let
$\predstyle{p}_{i-1}$ contain the rules

\begin{align*}
\predstyle{f}(x) & \to \colorpred_{?}(x)\\
\colorpred_{!}(x) \wedge \predstyle{f}(x) \wedge \predstyle{t}(y)& \to \colorpred_{?}(y)\\
\colorpred_{!}(x) \wedge \predstyle{t}(x) & \to \checkpred\\
\colorpred_{?}(x) \wedge
\predstyle{p}_{i}(\flagconst_1,\ldots,\flagconst_{i-1},x) & \to
\colorpred_{!}(x)\\
\end{align*}

Note that $\predstyle{p}_{0}$ is a Boolean \nlinmq query the size of
which is polynomial in the size of the input QBF.

Now, let $D$ be the database containing the two individuals $0$ and
$1$ as well as the facts $\predstyle{f}(0)$ and $\predstyle{t}(1)$.
We now show that the considered QBF is true exactly if $D\models
\predstyle{p}_{0}()$. To this end, we first note that by
construction the extension of $\predstyle{p}_L$ contains exactly
those $n$-tuples $\tuple{\delta_1,\ldots,\delta_n}$ for which the
corresponding truth value assignment $val$, sending $x_i$ to
$\mathbf{true}$ iff $\delta_i = 1$, makes the formula
$\bigwedge_{\ell\in L} \ell$ true. In the same way, the extension of
$\predstyle{p}_\mathcal{L}$ represents the set of truth value
assignments satisfying $\bigvee_{L\in\mathcal{L}} \bigwedge_{\ell\in
L} \ell$. Then, by descending induction, we can show that the
extensions of $\predstyle{p}_i$ encode the assignments to free
propositional variables of the subformula $Q_{i+1} x_{i+1} \ldots
Q_n x_n \bigvee_{L\in\mathcal{L}} \bigwedge_{\ell\in L} \ell$ that
make this formula true. Consequently, $\predstyle{p}_0$ has a
nonempty extension if the entire considered QBF is true.

Finally, the \NLogSpace-completeness for data complexity is again immediate,
where the upper bound is obtained from \lindatalog, and the lower bound follows from
the well-known hardness of reachability queries, which can be expressed in
\linmdatalogconst.
\end{proof}

\section*{Proofs for Section~\ref{sec_gqcontainmentup}}

\begin{repproposition}{prop_ruleauto}
There is an automaton $\aautomaton_{P,\arule}$ that accepts exactly the annotated matching trees for $\arule$ and $\aprogram$,
and which is exponential in the size of $\arule$ and $\aprogram$.
\end{repproposition}
\begin{proof}
We first construct an automaton $\aautomaton'_{P,\arule}$ that accepts matching trees where each node is
additionally annotated by a partial mapping of the form
$\Var(\arule)\to\mathcal{V}_{\aprogram}$ (called \emph{$\Var(\arule)$-label}), such that:
every special variable $x\in\Var(\arule)$ occurs in at least one $\Var(\arule)$-label,
and whenever a variable $x\in\Var(\arule)$ occurs in two, it is
mapped to the same variable and both variable occurrences are connected.
Note that this is essentially the same condition that we imposed for $\vec{\flagconst}$-annotations.

The intersection of tree automata can be computed in polynomial time.
We can therefore construct automata to check part of the conditions for (annotated) matching trees
to simplify the definitions.
We first construct an automaton $\aautomaton_x$ for checking the condition on $\Var(\arule)$-labels
for one variable $x\in\Var(\arule)$.
We define $\aautomaton_x=\tuple{\Sigma,Q_x,Q^s_x,\delta_x,Q^e_x}$, where the alphabet $\Sigma$ consists
of quadruples of proof-tree labels (from $\mathcal{R}_{\aprogram}$), $\vec{\flagconst}$-labels,
$p$-labels, and $\Var(\arule)$-labels.
The state set $Q_x$ is $\{a,b,\textsf{accept}\}\cup\{q_v\mid v\in\mathcal{V}_{\aprogram}\}$, signifying that
the current node is \emph{a}bove the first node annotated with a mapping for $x$, \emph{b}elow
or \emph{b}esides any nodes that were annotated with a mapping for $x$,
or at a node where $x$ is mapped to a variable $v$. That start-state set is $Q^s_x=\{a\}\cup\{q_v\mid v\in\mathcal{V}_{\aprogram}\}$; the end-state set if $Q^e_x=\{\textsf{accept}\}$.

Consider a rule $\arule'\in \mathcal{R}_{\aprogram}$
of the form $r_1(\vec{v_1})\wedge\ldots\wedge r_n(\vec{v_n})\wedge h_1(\vec{w_1})\wedge\ldots\wedge h_m(\vec{w_m}) \to h(\vec{v})$,
where $r_i$ are EDB predicates and $h_{(i)}$ are IDB predicates.
For the case that $m>0$, there is a transition $\tuple{q_1,\ldots,q_m}\in\delta(q, \tuple{\arule',\_,\_,\nu})$
exactly if the following conditions are satisfied:
\begin{itemize}
\item if $q=a$ and $\nu(x)$ is undefined, then $q_i=a$ for one $1\leq i\leq m$ and $q_j=b$ for all $1\leq j\leq m$ with $i\neq j$;
\item if $q=q_v$ and $\nu(x)=v$, then $q_i=q_v$ for all $1\leq i\leq m$ such that $v$ occurs in $\vec{w_i}$ and $q_i=b$ for all other $i$;
\item if $q=b$ and $\nu(x)$ is undefined, then $q_i=b$ for all $1\leq i\leq m$.
\end{itemize}
For the case $m=0$, there is a transition $\tuple{\textsf{accept}}\in\delta(q, \tuple{\arule',\_,\_,\nu})$
exactly if:
\begin{itemize}
\item if $q=q_v$ and $\nu(x)=v$;
\item if $q=b$ and $\nu(x)$ is undefined.
\end{itemize}
It is easy to check that the automaton $\aautomaton_x$ satisfies the required condition.
Now an automaton for checking the condition on $\Var(\arule)$-labels can be constructed
as the intersection $\aautomaton'_{\Var(\arule)}=\bigcap_{x\in\Var(\arule)}\aautomaton_x$. The automaton $\aautomaton'_{\vec{\flagconst}}$ for checking the
condition on $\vec{\flagconst}$-labels is constructed in a similar fashion.
Likewise, an automaton $\aautomaton'_p$ for checking the condition on $p$-labels is easy to define.

It remains to construct an automaton for checking the conditions (a)--(d) of Definition~\ref{defn_annotree}.
To do this, we interpret the $\Var(\arule)$-labels and $\vec{\flagconst}$-labels as partial
specifications of the required mapping $\nu$.
Condition (a) further requires that $\nu(\vec{x})=\vec{v}$, i.e., that the $\Var(\arule)$-label
at the unique node annotated with $p(\vec{v})$ contains this mapping. It is easy to
verify this with an automaton $\aautomaton'_{(a)}$.
Together, $\aautomaton'_{(a)}$, $\aautomaton'_{\vec{\flagconst}}$, and $\aautomaton'_{\Var(\arule)}$ provide a consistent variable
mapping that respects the $p$-label (a) and the connectedness of variable occurrences, i.e., (c) and (d).
To check the remaining condition (b), we use an automaton $\aautomaton'_{(b)}$.

The automaton for (b) will use auxiliary markers to record which atoms have been matched
in the current node and how exactly this was done. We record such a match
as a partial function from atoms $q(\vec{z})\in\varphi$ to instances $q(\vec{w})$
of such atoms using variables $\vec{w}\subseteq\mathcal{V}_{\aprogram}$.
The set of all such partial functions is denoted $\mathsf{Match}_{\varphi,\aprogram}$.
Note that this set is exponential (not double exponential).

We now define $\aautomaton'_{(b)}=\tuple{\Sigma,Q,Q_s,\delta,Q_e}$ where $\Sigma$ is as for $\aautomaton_x$ above.
The set of states $Q$ is $\{\textsf{accept}\}\cup (2^\varphi\times \mathsf{Match}_{\varphi,\aprogram})$,
where elements from $2^\varphi$ encode the subset of $\varphi$ that should be witnessed at or below the current node,
and the elements from $\mathsf{Match}_{\varphi,\aprogram}$ encode atoms that must be matched at the current node with their respective instantiations.
The start-state set $Q_s$ is $\{\tuple{\varphi,\mu}\mid \mu\in\mathsf{Match}_{\varphi,\aprogram}\}$; the end-state set $Q_e$ is $\{\textsf{accept}\}$.
The transition function $\delta$ is defined as follows. Consider a rule $\arule'\in \mathcal{R}_{\aprogram}$
of the form $r_1(\vec{v_1})\wedge\ldots\wedge r_n(\vec{v_n})\wedge h_1(\vec{w_1})\wedge\ldots\wedge h_m(\vec{w_m}) \to h(\vec{v})$,
where $r_i$ are EDB predicates and $h_{(i)}$ are IDB predicates.
For the case $m>0$,
there is a transition $\tuple{\tuple{\beta_1,\mu_1},\ldots,\tuple{\beta_m,\mu_n}}\in\delta( \tuple{\beta,\mu}, \tuple{\arule',\nu_{\vec{\flagconst}},\_,\nu_{\Var(\arule)}})$
exactly if the set $\beta\subseteq\varphi$ can be partitioned into sets $\beta',\beta_1,\ldots,\beta_m$ such that
$(\nu_{\vec{\flagconst}}\cup \nu_{\Var(\arule)})(\beta')=\mu(\beta')$ and $\mu(\beta')\subseteq\{r_1(\vec{v_1}),\ldots, r_n(\vec{v_n})\}$.
The element $\mu_i$ of successor states can be chosen freely; the validity of the choice will be checked later.
For the case $m=0$, there is a transition $\tuple{\textsf{accept}}\in\delta( \tuple{\beta,\mu}, \tuple{\arule',\nu_{\vec{\flagconst}},\_,\nu_{\Var(\arule)}})$
exactly if  $(\nu_{\vec{\flagconst}}\cup \nu_{\Var(\arule)})(\beta)=\mu(\beta)$ and $\mu(\beta)\subseteq \{r_1(\vec{v_1}),\ldots, r_n(\vec{v_n})\}$.
In fact, the information from $\mathsf{Match}_{\varphi,\aprogram}$ is not strictly necessary to define the transition,
since the relevant elements $\mu$ are always determined by other choices in the transition.
However, having this information explicit will be important in later proofs.

The automaton $\aautomaton'_{P,\arule}$ is obtained as the intersection
$\aautomaton'_{\Var(\arule)}\cap \aautomaton'_{\vec{\flagconst}}\cap \aautomaton'_p\cap \aautomaton'_{(a)}\cap \aautomaton'_{(b)}$.
It is easy to verify that it accepts exactly the $\Var(\arule)$-annotated matching trees.
Note that $\aautomaton'_{P,\arule}$ is exponential in size, already due to the exponentially large alphabet $\Sigma$.
Now the required automaton $\aautomaton_{P,\arule}$ is obtained by ``forgetting'' the $\Var(\arule)$-label in transitions of
$\aautomaton'_{P,\arule}$. This projection operation for tree automata is possible with a polynomial increase in size:
every state of $\aautomaton_{P,\arule}$ is a pair of a state of $\aautomaton'_{P,\arule}$ and a $\Var(\arule)$-label;
transitions of $\aautomaton_{P,\arule}$ are defined as for $\aautomaton'_{P,\arule}$, but keeping $\Var(\arule)$-label information
in states and introducing transitions for all possible $\Var(\arule)$-labels in child nodes.
\end{proof}

\begin{repproposition}{prop_ruleautoplus}
There is an alternating 2-way tree automaton $\aautomaton^+_{P,\arule,\vec{v}}$ that is polynomial in the size of
$\aautomaton_{P,\arule}$ such that, whenever $\aautomaton_{P,\arule}$ accepts a matching tree $T$ that has the $p$-annotation
$p(\vec{v})$ on node $e$, then $\aautomaton^+_{P,\arule,\vec{v}}$ has an accepting run that starts from the corresponding
node $e'$ on the tree $T'$ that is obtained by removing the $p$-annotation from $T$.
\end{repproposition}
\begin{proof}
Using alternating 2-way automata, we can traverse a tree starting from any node,
visiting each node once. To control the direction of the traversal, we create
multiple copies of each state $q$:
states $q_{\mathsf{down}}$ are processed like normal states in $\aautomaton_{P,\arule}$,
states $q_{\mathsf{up}}$ use an inverted transition of $\aautomaton_{P,\arule}$ to
move up the tree into a state $q_{\sigma,i}$; these auxiliary states are used to check
that the label of the upper node is actually $\sigma$ and to start new downwards processes
for all child nodes other than the one ($i$) that we came from.

To ensure that the constructed automaton $\aautomaton^+_{P,\arule,\vec{v}}$ simulates the behavior of
$\aautomaton_{P,\arule}$ in case the annotation $p(\vec{v})$ is found, we eliminate all transitions
that mention other $p$-annotations.
Moreover, we assume without loss of generality
that the states of $\aautomaton_{P,\arule}$ that allow a transition mentioning $p(\vec{v})$
cannot be left through any other transition; this can always be ensured by
duplicating states and using them exclusively for one kind of transition.
Let $Q_p$ be the set of states of $\aautomaton_{P,\arule}$ that admit (only) transitions
mentioning $p(\vec{v})$.
Let $\aautomaton'_{P,\arule}=\tuple{\Sigma',Q,Q_s,\delta',Q_e}$ denote the automaton
over the alphabet $\Sigma'$ of $\vec{\flagconst}$-annotated proof trees (without $p$-annotations),
with the same (start/end) states as $\aautomaton_{P,\arule}$, and where $\delta'$ is defined based on the transition function $\delta$ of $\aautomaton_{P,\arule}$ as follows: $\delta'(\tuple{\arule',M})$
is the union of all sets of the form $\delta(\tuple{\arule',\vec{\flagconst}\text{-label},p\text{-label}})$
where $p\text{-label}$ is either $p(\vec{v})$ or empty.
By this construction, there is a correspondence between the accepting runs of $\aautomaton_{P,\arule}$ over trees
where one node $e$ is annotated with $p(\vec{v})$ and accepting runs of $\aautomaton'_{P,\arule}$ (on trees without $p$-annotations) for which the node $e$ is visited in some state of $Q_p$.

%

%

Let $s$ be the maximal out-degree of
proof trees for $P$, i.e., the maximal number of IDB atoms in bodies of $P$.
The state set $Q^+$ of $\aautomaton^+_{P,\arule,\vec{v}}$ is given by the disjoint union
$\{q_{\mathsf{up}}\mid q\in Q\}\cup\{q_{\sigma,i}\mid q\in Q,\sigma\in\Sigma, 1\leq i\leq s\}\cup\{q_{\mathsf{down}}\mid q\in Q\}\cup\{\mathsf{start},\mathsf{accept}\}$.
The start-state set is $Q^+_s=\{\mathsf{start}\}$ and the end-state set is $Q^+_e=\{\mathsf{accept}\}\cup\{q_{\mathsf{down}}\mid q\in Q_e\}$.

Transitions of $\aautomaton^+_{P,\arule,\vec{v}}$ are defined as follows:
\begin{itemize}
\item For all $\sigma\in\Sigma$, let $\delta^+(\mathsf{start},\sigma)$ be the disjunction of all formulae $\tuple{0,q_{\mathsf{up}}}\wedge\tuple{0,q_{\mathsf{down}}}$ where $q\in Q_p$.
\item For states $q_{\mathsf{down}}$ and $\sigma\in\Sigma$, let $\delta^+(q_{\mathsf{down}},\sigma)$ be the disjunction of all formulae $\tuple{1,q^1_{\mathsf{down}}}\wedge\ldots\wedge\tuple{m,q^m_{\mathsf{down}}}$ for
which $\aautomaton'_{P,\arule}$ has a transition $\tuple{q^1,\ldots,q^m}\in\delta'(q,\sigma)$.
\item For states $q_{\mathsf{up}}$ and $\sigma\in\Sigma$, let $\delta^+(q_{\mathsf{up}},\sigma)$
be the disjunction of all formulae $\tuple{-1,q'_{\sigma',i}}$
for which $\aautomaton'_{P,\arule}$ has a transition $\tuple{q^1,\ldots,q^{i-1},q,q^{i+1},\ldots,q^m}\in\delta'(q',\sigma')$
and the current node is the $i$th child of its parent (we can assume that this information is encoded in the labels $\sigma$, even for basic proof trees, which increases the alphabet only linearly; we omit this in our definitions since it would clutter all other parts of our proof without need).
\item For states $q_{\sigma,i,q'}$, let $\delta^+(q_{\sigma,i,q'},\sigma)$ be the disjunction
of all formulae $\tuple{0,q_{\mathsf{up}}}\wedge\tuple{1,q^1_{\mathsf{down}}}\wedge\ldots\wedge\tuple{i-1,q^{i-1}_{\mathsf{down}}}\wedge\tuple{i+1,q^{i+1}_{\mathsf{down}}}\wedge\ldots\wedge\tuple{m,q^m_{\mathsf{down}}}$ for which $\aautomaton'_{P,\arule}$ has a transition $\tuple{q^1,\ldots,q^{i-1},q',q^{i+1},q^m}\in\delta'(q,\sigma)$.
\item For all starting states $q\in Q_s$ of $\aautomaton'_{P,\arule}$ and $\sigma\in\Sigma$, let $\delta(q_{\mathsf{up}},\sigma)=\tuple{0,\mathsf{accept}}$.
\end{itemize}
It is not hard to verify that $\aautomaton^+_{P,\arule,\vec{v}}$ has the required properties.
\end{proof}

\begin{repproposition}{prop_containmentaltauto}
For a \datalog query $P$ and a \gq query $P'$ with special constants $\vec{\flagconst}$, there is an alternating 2-way automaton
$\aautomaton^+_{P\sqsubseteq P'}$ of exponential size that accepts
the $\vec{\flagconst}$-annotated proof trees of $P$ that encode expansion trees with $\vec{\flagconst}$ assignments for
which $P'$ has a match.
\end{repproposition}
\begin{proof}
Let $P'$ be the set $\{\arule_1,\ldots,\arule_\ell\}$.
For every IDB predicate $p$, let $P'_p$ denote the set of rules in $P'$ with head predicate $p$ (possibly $\checkpred$).
Without loss of generality, we assume that distinct rules use distinct sets of variables.
For every frontier-guarded rule $\arule'$, let $\mathsf{guard}(\arule')$ be a fixed EDB atom that acts as a guard
in this rule, i.e., an atom that refers to all variables in the head of $\arule'$.

Consider a rule $\arule'\in P'$ with IDB atoms $q_1(\vec{t_1}),\ldots,q_m(\vec{t_m})$ in its body.
We construct new rules from $\arule'$ by replacing each atom $q_i(\vec{t_i})$ with
a guard atom $\mathsf{guard}(\arule'_i)$, suitably unified.
Formally, assume that there are rules $\arule'_i\in P'_{q_i}$ with head $q_i(\vec{s_i})$ and a substitution $\theta$
that is a most general unifier for the problems $\vec{t_i}\theta = \vec{s_i}\theta$, for all $i\in\{1,\ldots,m\}$,
and that maps every variable in $\arule'_i$ that does not occur in the head to a globally fresh variable.
Then the \emph{guard expansion} of $\arule'$ for $(\arule'_i)_{i=1}^m$ and $\theta$ is
the rule that is obtained from $\arule'\theta$ by replacing each body atom $q_i(\vec{t_i})\theta$
by $\mathsf{guard}(\arule'_i)\theta$. By construction, two distinct atoms $\mathsf{guard}(\arule'_i)\theta$
and $\mathsf{guard}(\arule'_j)\theta$ do not share variables, unless at positions that correspond to head variables
in rules $\arule'_i$ and $\arule'_j$.
The atoms $\mathsf{guard}(\arule'_i)\theta$ in a guard expansion are called \emph{replacement guards}.
We consider two guard expansions to be equivalent if they only differ in the choice of the most general unifier.
Let $\mathsf{Guard}(\arule')$ be the set of all guard expansions of $\arule'\in P'$,
i.e., a set containing one representative of each class of equivalent guard expansions.
$\mathsf{Guard}(\arule')$ is exponential since there are up to $|P'|^m$ non-equivalent guard expansions for a rule with $m$ IDB atoms.

The automaton $\aautomaton^+_{P\sqsubseteq P'}$ is constructed as follows.
For every guard expansion $\arule_g\in \bigcup_{\arule'\in P'}\mathsf{Guard}(\arule')$
and every list $\vec{v}$ of proof-tree variables of the arity of the head of $\arule_g$,
consider the alternating 2-way tree automaton $\aautomaton^+_{P,\arule_g,\vec{v}}$
of Proposition~\ref{prop_ruleautoplus}. We assume w.l.o.g.\ that the state sets of these automata
are mutually disjoint.
Let $\aautomaton^+_{P\sqsubseteq P'}=\tuple{\Sigma,Q,Q_s,\delta,Q_e}$. As before, $\Sigma$
consists of pairs of a rule instance from $\mathcal{R}_{\aprogram}$ and a partial mapping of $\vec{\flagconst}$
to $\mathcal{V}_{\aprogram}$.
The state set $Q$ is the disjoint union of all state sets of the automata of form $\aautomaton^+_{P,\arule_g,\vec{v}}$.
The start-state set $Q_s$ is the disjoint union of all start-state sets of automata $\aautomaton^+_{P,\arule_g,\vec{v}}$
for which $\arule_g$ is a guard expansion of a rule with head $\checkpred$ (and $\vec{v}$ is the empty list).
The end-state set $Q_e$ is the disjoint union of all end-state sets of automata $\aautomaton^+_{P,\arule_g,\vec{v}}$.

The transition function $\delta$ is defined as follows.
By the construction in Proposition~\ref{prop_ruleauto}, each state $q$ in the automaton
$\aautomaton_{P,\arule}$ encodes a partial mapping $\mathsf{match}(q)$ from body atoms of $\arule$ to
instantiated atoms that use variables from $\mathcal{V}_{\aprogram}$, which are matched at the current tree node.
This information is preserved through alphabet projections, intersections, and even through the construction
in Proposition~\ref{prop_ruleautoplus}. We can therefore assume that each state $q$ of $\aautomaton^+_{P\sqsubseteq P'}$
is associated with a partial mapping $\mathsf{match}(q)$.

For every state $q\in Q_{P,\arule_g,\vec{v}}$ and every $\sigma\in\Sigma$, we define
$\delta(q,\sigma) = \delta_{P,\arule_g,\vec{v}}(q,\sigma)\wedge \psi$, where
$\psi$ defined as follows. For every replacement guard atom $\alpha$ of $\arule_g$
for which $\mathsf{match}(q)(\alpha)$ is defined, we consider the formula
$\psi_\alpha=\tuple{0,q_1}\vee\ldots\vee\tuple{0,q_\ell}$, where
\begin{itemize}
\item $\alpha=\mathsf{guard}(\arule')\theta$ for some rule $\arule'$ and substitution $\theta$;
\item $\mathsf{match}(q)(\alpha)=\alpha\theta'$ for some substitution $\theta'$;
\item $q_1,\ldots q_\ell$ are the start states of the automaton $\aautomaton_{P,\arule',\vec{z}\theta\theta'}$ where $p(\vec{z})$ is the head of $\arule'$.
\end{itemize}
Now $\psi$ is the conjunction of all formulae $\psi_\alpha$ thus defined.
%
\end{proof}

\section*{Proofs for Section~\ref{sec_atmencoding}}

\begin{figure*}[t!]
\[
\begin{array}{r@{~~~}l}
\multicolumn{2}{l}{\textbf{(1)~~Unique head marker and correct left/right head markers:}}\\[0.5ex]
\textsf{Head}(y,p_1)\wedge\textsf{NextCell}(y,z)\wedge\textsf{Head}(z,p_2) & \text{where $\tuple{p_1,p_2}\in\{\tuple{h,h},\tuple{h,l},\tuple{r,h},\tuple{r,l}\}$}\\[0.5ex]
\textsf{Head}(y,h)\wedge\textsf{Head}(y,p) & \text{where $p\in\{r,l\}$}\\[1ex]
\multicolumn{2}{l}{\textbf{(2)~~Unique start configuration:}}\\[0.5ex]
\textsf{FirstConf}(x,y)\wedge\textsf{State}_q(y) & \text{where $q\neq q_s$}\\[0.5ex]
\textsf{FirstConf}(x,y)\wedge\textsf{FirstCell}(y,z)\wedge\textsf{Head}(z,p)& \text{where $p\in\{l,r\}$}\\[0.5ex]
\textsf{FirstConf}(x,y)\wedge\textsf{ConfCell}(y,z)\wedge\textsf{Symbol}(z,c_\sigma) & \text{where $\sigma\neq \square$}\\[1ex]
\multicolumn{2}{l}{\textbf{(3)~~Valid, uniquely defined transitions:}}\\[0.5ex]
\textsf{State}_q(y)\wedge \textsf{Head}(z,h)\wedge\textsf{ConfCell}(y,z)\wedge\textsf{Symbol}(z,c_\sigma)\wedge \textsf{NextConf}_\delta(y,y')\wedge{}
	&\text{where $\delta=\tuple{q_1,\sigma_1,q_2,\sigma_2,d}$}\\
\textsf{State}_{q'}(y')\wedge\textsf{ConfCell}(y',z')\wedge\textsf{SameCell}(z',z)\wedge\textsf{Symbol}(z',c_{\sigma'})
	&\text{with $q_1\neq q$ or $\sigma_1\neq\sigma$ or $q_2\neq q'$ or $\sigma_2\neq\sigma'$}\\[1ex]
\multicolumn{2}{l}{\textbf{(4)~~Unique end state:}}\\[0.5ex]
\textsf{LastConf}(y)\wedge\textsf{State}_q(y) & \text{where  $q\neq q_e$}\\[1ex]
\multicolumn{2}{l}{\textbf{(5)~~Memory:}}\\[0.5ex]
\textsf{ConfCell}(y_1,x_1)\wedge\textsf{Head}(x_1,r)\wedge\textsf{Symbol}(x_1,c_\sigma)\wedge\textsf{NextConf}_\delta(y_1,y_2)\wedge{} &\\
\textsf{ConfCell}(y_2,x_2)\wedge\textsf{SameCell}(x_1,x_2)\wedge\textsf{Symbol}(x_2,c_{\sigma'})
	&\text{where $\sigma\neq\sigma'$}\\[0.5ex]
\textsf{ConfCell}(y_1,x_1)\wedge\textsf{Head}(x_1,l)\wedge\textsf{Symbol}(x_1,c_\sigma)\wedge\textsf{NextConf}_\delta(y_1,y_2)\wedge{} &\\
\textsf{ConfCell}(y_2,x_2)\wedge\textsf{SameCell}(x_1,x_2)\wedge\textsf{Symbol}(x_2,c_{\sigma'})
	&\text{where $\sigma\neq\sigma'$}\\[1ex]
\multicolumn{2}{l}{\textbf{(6)~~Head movement:}}\\[0.5ex]
\textsf{ConfCell}(y_1,x_1)\wedge\textsf{Head}(x_1,h)\wedge\textsf{NextConf}_\delta(y_1,y_2)\wedge{}
	&\text{where $\delta=\tuple{q_1,\sigma_1,q_2,\sigma_2,\text{right}}$}\\
\textsf{ConfCell}(y_2,x_2)\wedge\textsf{SameCell}(x_1,x_2)\wedge\textsf{NextCell}(x_2,x_2')\wedge\textsf{Head}(x_2',p)
	&\text{and $p\in\{r,l\}$}\\[0.5ex]
\textsf{ConfCell}(y_1,x_1)\wedge\textsf{Head}(x_1,h)\wedge\textsf{NextConf}_\delta(y_1,y_2)\wedge{}
	&\text{where $\delta=\tuple{q_1,\sigma_1,q_2,\sigma_2,\text{right}}$}\\
\textsf{ConfCell}(y_2,x_2)\wedge\textsf{SameCell}(x_1,x_2)\wedge\textsf{LastCell}(x_2)\wedge\textsf{Head}(x_2,p)
	&\text{and $p\in\{r,l\}$}\\[0.5ex]
\textsf{ConfCell}(y_1,x_1)\wedge\textsf{Head}(x_1,h)\wedge\textsf{NextConf}_\delta(y_1,y_2)\wedge{}
	&\text{where $\delta=\tuple{q_1,\sigma_1,q_2,\sigma_2,\text{left}}$}\\
\textsf{ConfCell}(y_2,x_2)\wedge\textsf{SameCell}(x_1,x_2)\wedge\textsf{NextCell}(x_2',x_2)\wedge\textsf{Head}(x_2',p)
	&\text{and $p\in\{r,l\}$}\\[0.5ex]
\textsf{ConfCell}(y_1,x_1)\wedge\textsf{Head}(x_1,h)\wedge\textsf{NextConf}_\delta(y_1,y_2)\wedge{}
	&\text{where $\delta=\tuple{q_1,\sigma_1,q_2,\sigma_2,\text{left}}$}\\
\textsf{ConfCell}(y_2,x_2)\wedge\textsf{SameCell}(x_1,x_2)\wedge\textsf{FirstCell}(z,x_2)\wedge\textsf{Head}(x_2,p)
	&\text{and $p\in\{r,l\}$}\\[0.5ex]
\end{array}
\]
\caption{Queries to construct a containment encoding as in Lemma~\ref{lemma_atmquasienctoenc}\label{fig_atmquasienctoenc}}
\end{figure*}
\begin{replemma}{lemma_atmquasienctoenc}
Consider an ATM $\mathcal{M}$, and queries as in Definition~\ref{def_atmencode}, including $\textsf{SameCell}[x,y]$,
that are \kmq{k} queries for some $k\geq 0$.
There is a \kmq{k} query $P[x]$, polynomial in the size of $\mathcal{M}$ and the given queries, such that the following hold.
\begin{itemize}
\item For every accepting run of $\mathcal{M}$ in space $s$, there is some database instance $\Inter$ with some element $c$ that encodes the run, such that $c\notin P^\Inter$.
\item If an element $c$ of $\Inter$ encodes a tree of quasi-configurations of $\mathcal{M}$ in space $s$, and if $c\notin P^\Inter$,
then $c$ encodes an accepting run of $\mathcal{M}$ in space $s$.
\end{itemize}
Moreover, if all input queries are in \klinmq{k}, then so is $P$.
\end{replemma}
\begin{proof}
We construct $P$ from all (polynomially many) positive queries obtained by instantiating the query patterns in Figure~\ref{fig_atmquasienctoenc}.
Since $P$ needs to be a unary query with variable $x$, we extend every positive query that does not contain $x$
with the atom $\textsf{FirstConf}[x,x']$ (omitted for space reasons in Figure~\ref{fig_atmquasienctoenc}).
By Proposition~\ref{prop_posfcqs} we can express the disjunctions of all the positive queries in Figure~\ref{fig_atmquasienctoenc}
as a \klinmq{k} $P[x]$ of polynomial size (for $k=0$ it is a \ucq).

If an element $c$ in a database instance $\Inter$ encodes an accepting run of $\mathcal{M}$ in space $s$, and $\Inter$ contains no other structures,
then none of the queries in Figure~\ref{fig_atmquasienctoenc} matches. Hence $c\notin P^\Inter$.

Conversely, assume that $c$ encodes a tree of $\mathcal{M}$ quasi-configurations in space $s$ and $c\notin P^\Inter$.
If none of the queries in Figure~\ref{fig_atmquasienctoenc} (1) match, the head positions of
every configuration must form a sequence $l,\ldots,l,h,r,\ldots,r$;
hence all quasi-configurations are actually configurations.
Queries (2)--(4) ensure that the first and last configuration are in the start and end state, respectively, and that each transition
is matched by suitable state and tape modifications. Queries (5) ensure that tape cells that are not at the head of the TM are not modified
between configurations. Queries (6) ensure that the movement of the head is consistent with the transitions, and especially does not leave the prescribed space.
Note that the queries allow transitions that try to move the head beyond the tape and require that the head stays in its current position in this case. This
allows the ATM to recognize the end of the tape, which is important for the Turing machines that we consider below.
With all these restrictions observed, $c$ must encode a run of $\mathcal{M}$ in space $s$.
\end{proof}

\section*{Proofs for Section~\ref{sec_mqcontainmentlow}}

\begin{replemma}{lemma_mducqexpatm}
For any ATM $\mathcal{M}$, there is an \mdatalogconst query $P_1[x]$, a \linmq $P_2[x]$, queries as in Definition~\ref{def_atmencode}
that are \linmq{}s, and a same-cell query that is a \ucq, such that $P_1[x]$ and $P_2[x]$ containment-encode accepting
runs of $\mathcal{M}$ in exponential space.
\end{replemma}
\begin{proof}
Let $\mathcal{M}=\tuple{Q,\Sigma,\Delta,q_s,q_e}$ with $Q$
partitioned into existential states $Q_\exists$ and universal states $Q_\forall$.
In order to use Lemma~\ref{lemma_atmquasienctoenc}, we first construct queries
$P'_1$ and $P'_2$ that containment-encode quasi-configuration trees of $\mathcal{M}$
in space $2^\ell$ for some $\ell$ that is linear in the size of the queries (w.r.t.\ to
suitable queries as in Definition~\ref{def_atmencode}).

Our signature contains the binary predicates (distinguished from the queries of Definition~\ref{def_atmencode} by
using lower case letters)
$\textsf{firstConf}$, $\textsf{nextConf}_\delta$ for all $\delta\in\Delta$,
$\textsf{firstCell}$, $\textsf{nextCell}$,
$\textsf{bit}_i$ for all $i\in\{1,\ldots,\ell\}$,
$\textsf{symbol}$, $\textsf{head}$,
as well as the unary predicates $\textsf{lastConf}$, and
$\textsf{state}_q$ for all $q\in Q$.

We define $P'_1$ to be the following \mdatalogconst query that has the goal predicate
$\colorpred_{\textit{goal}}$ and uses two further constants $0$ and $1$:

\noindent{\small%
\begin{align*}
\textsf{firstConf}(x,y)\wedge\colorpred_{\textit{conf}}(y) &\to\colorpred_{\textit{goal}}(x)\\
%
\textsf{state}_q(x)\wedge\textsf{firstCell}(x,y)\wedge\colorpred_{\textit{bit}_1}(y) &\to\colorpred_{\textit{conf}}(x) & \text{for $q\in Q$}\displaybreak[0]\\
\textsf{bit}_{i-1}(x,0)\wedge\colorpred_{\textit{bit}_i}(x) &\to\colorpred_{\textit{bit}_{i-1}}(x) & \text{for $i\in\{2,\ldots,\ell\}$}\displaybreak[0]\\
\textsf{bit}_{i-1}(x,1)\wedge\colorpred_{\textit{bit}_i}(x) &\to\colorpred_{\textit{bit}_{i-1}}(x) & \text{for $i\in\{2,\ldots,\ell\}$}\displaybreak[0]\\
\textsf{symbol}(x,c_\sigma)\wedge\colorpred_{\textit{symbol}}(x) &\to\colorpred_{\textit{bit}_\ell}(x) & \text{for $\sigma\in\Sigma$}\displaybreak[0]\\
\textsf{head}(x,h)\wedge\colorpred_{\textit{head}}(x) &\to\colorpred_{\textit{symbol}}(x) &\displaybreak[0]\\
\textsf{head}(x,l)\wedge\colorpred_{\textit{head}}(x) &\to\colorpred_{\textit{symbol}}(x) &\displaybreak[0]\\
\textsf{head}(x,r)\wedge\colorpred_{\textit{head}}(x) &\to\colorpred_{\textit{symbol}}(x) & \displaybreak[0]\\
%
%
\textsf{nextCell}(x,y)\wedge\colorpred_{\textit{bit}_1}(y) &\to\colorpred_{\textit{head}}(x) & \displaybreak[0]\\
\textsf{nextConf}_\delta(x,y)\wedge\colorpred_{\textit{conf}}(y) &\to\colorpred_{\textit{head}}(x) & \text{for $\delta=\tuple{q,\sigma,q',\sigma',d}$}\\[-0.7ex]
 && \text{with $q\in Q_\exists$}\displaybreak[0]\\
%
\textsf{nextConf}_{\delta_1}(x,y_1)\wedge\colorpred_{\textit{conf}}(y_1) \wedge{}&  &\text{for $\delta_1=\tuple{q,\sigma,q',\sigma',d}$, }\\[-0.7ex]
\textsf{nextConf}_{\delta_2}(x,y_2)\wedge\colorpred_{\textit{conf}}(y_2) &\to\colorpred_{\textit{head}}(x) & \text{$q\in Q_\forall$, and $\delta_1\neq\delta_2$} \displaybreak[0]\\
\textsf{lastConf}(x) &\to\colorpred_{\textit{head}}(x) & \displaybreak[0]\\
\end{align*}%
}%
$P'_1$ encodes structures that resemble configuration trees, but with each configuration ``tape''
consisting of an arbitrary sequence of ``cells'' of the form
$\textsf{bit}_1(x,v_1),\ldots,\textsf{bit}_\ell(x,v_\ell),\textsf{symbol}(x,c_\sigma),\textsf{head}(x,p)$,
where each $v_i$ is either $0$ or $1$.
The values for the bit sequence encode a binary number of length $\ell$.
We provide a query $P'_2$ which ensures that each sequence of cells
encodes an ascending sequence of binary numbers from $00\ldots0$ to $11\ldots1$.
More precisely, $P'_2$ checks if there are any consecutive cells that
violate this rule, i.e., the structures matched by $P'_1$ but not by $P'_2$
are those where each configuration contains $2^\ell$ cells.
The following query checks whether bit $i$ is the rightmost bit containing a $0$ and
bit $i$ in the successor configuration also contains a $0$, which is a situation that must not
occur if the bit sequences encode a binary counter:

\noindent{\small%
\begin{align*}
\textsf{bit}_i(y,0)\wedge\textsf{bit}_{i+1}(y,1)\wedge\ldots\wedge\textsf{bit}_\ell(y,1)\wedge
 \textsf{nextCell}(y,z)\wedge\textsf{bit}_i(z,0)
\end{align*}%
}%

In a similar way, we can ensure that every bit to the right of the rightmost $0$ is changed to
$0$, every bit that is left of a $0$ remains unchanged,
the first number is $0\ldots0$, and the last number is $1\ldots1$. The query $P'_2$ is the
union of all of these (polynomially many) conditions, each with new atom
$\textsf{firstConf(x,y)}$ added and all variables other than $x$ existentially quantified; this ensures
that we obtain a unary query that matches the same elements as $P'_1$ if it matches at all.

We claim that the elements matching $P'_1$ but not $P'_2$ encode quasi-configuration
trees of $\mathcal{M}$ in space $2^\ell$. Indeed, it is easy to specify the queries required by Definition~\ref{def_atmencode}.
The most complicated query is $\textsf{ConfCell}[x,y]$, which can be defined by the following \linmq{}:

\noindent{\small%
\begin{align*}
\textsf{state}_q(\flagconst_1)\wedge\textsf{nextCell}(\flagconst_1,y) &\to \colorpred(y) & \text{for all $q\in Q$}\\
\colorpred(y)\wedge\textsf{nextCell}(y,z) &\to \colorpred(y)\\
\colorpred(\flagconst_2) &\to\checkpred
\end{align*}%
}%

\noindent The remaining queries are now easy to specify, where we use $\textsf{ConfCell}[x,y]$, knowing that
a conjunctive query over \linmq{}s can be transformed into a single \linmq{} using Proposition~\ref{prop_posfcqs}:

\noindent{\small%
\begin{align*}
\textsf{FirstConf}[x,y] \defeq{}& \textsf{firstConf}(x,y)\\
\textsf{NextConf}_\delta[x,y] \defeq{}& \exists z.\textsf{ConfCell}(x,z)\wedge\textsf{nextConf}_{\delta}(z,y)\\
\textsf{LastConf}[x] \defeq{}& \exists z.\textsf{ConfCell}(x,z)\wedge\textsf{lastConf}(z)\displaybreak[0]\\
\textsf{State}_q[x] \defeq{}& \textsf{state}_q(x)\displaybreak[0]\\
\textsf{Head}[x,y] \defeq{}& \textsf{head}(x,y) \displaybreak[0]\\
\textsf{FirstCell}[x,y] \defeq{}& \textsf{firstCell}(x,y))\displaybreak[0]\\
\textsf{NextCell}[x,y] \defeq{}& \textsf{nextCell}(x,y)\displaybreak[0]\\
\textsf{LastCell}[x] \defeq{}& \textsf{lastConf}(x)\vee \exists z.\textsf{nextConf}(x,z)\displaybreak[0]\\
\textsf{Symbol}[x,y] \defeq{}&  \textsf{symbol}(x,y)\\
\textsf{SameCell}[x,y] \defeq{}& \exists v_1,\ldots,v_\ell.\textsf{bit}_1(x,v_1)\wedge\textsf{bit}_1(y,v_1)\wedge{}\\
	&\ldots\wedge\textsf{bit}_\ell(x,v_\ell)\wedge\textsf{bit}_\ell(y,v_\ell)
\end{align*}%
}%
Using these queries, we can construct a \linmq $P$ as in Lemma~\ref{lemma_atmquasienctoenc}
such that $P_1=P'_1$ and $P_2=P'_2 \vee P$ containment-encode
accepting runs of $\mathcal{M}$.
\end{proof}

\begin{replemma}{lemma_atmencodenesting}
Assume that there is some space bound $s$ such that, for every DTM $\mathcal{M}$, there is a
\mdatalogconst query $P_1[x]$ and an \kmq{k+1} query $P_2[x]$ with $k\geq 0$, such that
$P_1[x]$ and $P_2[x]$ containment-encode accepting runs of $\mathcal{M}$ in $s$,
where the queries required by Definition~\ref{def_atmencode} are \kmq{k+1} queries.
Moreover, assume that there is a suitable same-cell query that is in \kmq{k}.

Then, for every ATM $\mathcal{M}'$, there is a \mdatalogconst query $P_1'[x]$, an \kmq{k+1} $P_2'[x]$,
and \kmq{k+1} queries as in Definition~\ref{def_atmencode}, such that
$P_1'[x]$ and $P_2'[x]$ containment-encode an accepting run of $\mathcal{M}'$ in space $s'\geq 2^s$.
Moreover, the size of the queries for this encoding is polynomial in the size of the queries for the original encoding.
\end{replemma}
\begin{proof}
There is a TM $\mathcal{M}=\tuple{Q,\Sigma,\Delta,q_s,q_e}$ that counts from $0$ to $2^s$ in binary (using space $s$) and then halts.
$\mathcal{M}$ can be small (constant size) since our formalization of (A)TMs allows the TMs to recognize the
last tape position to ensure that the maximal available space is used.
The computation will necessarily take $s'>2^s$ steps to complete since multiple steps are needed to increment the counter by $1$.
Let $P_1[x]$ and $P_2[x]$ be queries that containment-encode accepting runs of $\mathcal{M}$ in $s$, and let
$\textsf{ConfCell}$, $\textsf{SameCell}$, etc.\ denote the respective \klinmq{k} as in Definition~\ref{def_atmencode}.

Let $\mathcal{M}'=\tuple{Q',\Sigma',\Delta',q_s',q_e'}$ be an arbitrary ATM.
We use the signature of $P_1$, extended by additional binary predicates
$\textsf{firstConf}'$, $\textsf{nextConf}'_\delta$ for all $\delta\in\Delta'$,
$\textsf{symbol}'$, $\textsf{head}'$,
as well as unary predicates $\textsf{lastConf}'$, and
$\textsf{state}'_q$ for all $q\in Q'$.
All of these are assumed to be distinct from predicates in $P_1$.

Let $\colorpred_\text{goal}$ be the goal predicate of $P_1$, and let
$\colorpred_\text{tape}$ be a new unary IDB predicate. We construct the program $\bar{P}_1$
from $P_1$ as follows.
For every rule of $P_1$ that does
not contain an IDB atom in its body we add the atom $\colorpred_\text{tape}(x)$ to the body,
where $x$ is any variable that occurs in the rule.
Intuitively speaking, the IDBs $\colorpred_\text{tape}$ and $\colorpred_\text{goal}$ mark the start and
end of tapes of $\mathcal{M}'$, which are represented by runs of $\mathcal{M}$.
Moreover, we modify $\bar{P}_1$ to ``inject'' additional state and head information for $\mathcal{M}'$ into configurations of $\mathcal{M}$,
i.e., we extend $P_1$ to ensure that every element $e$ with $\textsf{state}_q(e)$ also occurs in
some $\textsf{symbol}'(e,c'_{\sigma'})$ and in some relation $\textsf{head}'(e,p)$.
This can always be achieved by adding a linear number of IDB predicates and rules.

Now $P'_1$ is defined to be a \mdatalogconst query with goal predicate $\colorpred'_\text{goal}$
(assumed, like all IDB predicates of form $\colorpred'$ below, to be distinct from any IDB predicate in $\bar{P}_1$),
which is obtained as the union of $\bar{P}_1$ with the following rules:

\noindent{\small%
\begin{align*}
\textsf{firstConf}'(x,y)\wedge\colorpred'_{\textit{conf}}(y) &\to\colorpred'_{\textit{goal}}(x)\\
\textsf{state}'_q(x)\wedge\colorpred_{\textit{goal}}(x) &\to\colorpred'_{\textit{conf}}(x) & \text{for $q\in Q$}\displaybreak[0]\\
\textsf{nextCell}'(x,y)\wedge\colorpred_{\textit{goal}}(y) &\to\colorpred_{\textit{tape}}(x) & \text{for $q\in Q$}\displaybreak[0]\\
\textsf{nextConf}'_\delta(x,y)\wedge\colorpred'_{\textit{conf}}(y) &\to\colorpred_{\textit{tape}}(x) & \text{for $\delta=\tuple{q,\sigma,q',\sigma',d}$}\\[-0.7ex]
 && \text{with $q\in Q_\exists$}\displaybreak[0]\\
%
\textsf{nextConf}'_{\delta_1}(x,y_1)\wedge\colorpred'_{\textit{conf}}(y_1) \wedge{} & &\text{for $\delta_1=\tuple{q,\sigma,q',\sigma',d}$, }\\[-0.7ex]
\textsf{nextConf}'_{\delta_2}(x,y_2)\wedge\colorpred'_{\textit{conf}}(y_2) &\to\colorpred_{\textit{tape}}(x) & \text{$q\in Q_\forall$, and $\delta_1\neq\delta_2$} \displaybreak[0]\\
\textsf{lastConf}'(x) &\to\colorpred_{\textit{tape}}(x) & \displaybreak[0]\\
\end{align*}%
}%

\noindent
$P'_1$ encodes trees of trees of $\mathcal{M}$ quasi-configurations in space $s$.
The structures matched by $P'_1$ but not by $P_2$ encode trees of accepting runs of
$\mathcal{M}$ in space $s$ (note that these runs are linear, since $\mathcal{M}$ is not alternating).
Every such run consists of the same number $s'\geq 2^s$ of configurations;
these configurations represent the tape cells of our encoding of $\mathcal{M}'$ sequences.
This encoding is formalized by queries as follows.
The queries $\textsf{FirstConf}'[x,y]$, $\textsf{State}'_q[x]$, $\textsf{Head}'[x,y]$,
and $\textsf{Symbol}'[x,y]$
are directly expressed by singleton \cq{}s that use the eponymous predicates $\textsf{firstConf}'(x,y)$, etc.
To access cells of $\mathcal{M}'$, we can use the analogous queries to access configurations
of $\mathcal{M}$: $\textsf{FirstCell}'[x,y]=\textsf{FirstConf}(x,y)$,
$\textsf{NextCell}'[x,y]=\textsf{NextConf}(x,y)$, and
$\textsf{LastCell}'[x]=\textsf{LastConf}(x)$.

The remaining queries can be expressed as \linmq queries. To present these queries in a more readable
way, we specify them in regular expression syntax rather than giving many rules for each.
It is clear that regular expressions over unary and binary predicates can be expressed in \linmq
(it was already shown that \mq{}s can express regular path queries, which is closely related \cite{RK13:flagcheck}).
We use abbreviation $\textsf{P1SYMBOL}$ to express the regular expression that is a disjunction
of all predicate symbols that occur in $P_1$ (this allows us to skip over any structures generated
by $P_1$; with the specific forms of $P_1$ that can occur in our proofs, one could make this more specific
to use only certain binary predicates, but our formulation does not depend on internals of $P_1$).
Moreover, let $\textsf{STATE}$ be the disjunction of all
atoms $\textsf{state}'_q(x)$ and $\exists y.\textsf{head}'(x,y)$ (both unary).

\noindent{\small%
\begin{align*}
\textsf{NextConf}'_\delta[x,y] \defeq{}& \textsf{STATE}~\textsf{P1SYMBOL}^\ast~\textsf{nextConf}'_{\delta}\\
\textsf{LastConf}'[x] \defeq{}& \textsf{STATE}~\textsf{P1SYMBOL}^\ast~\textsf{lastConf}'\\
\textsf{ConfCell}'[x,y] \defeq{}& \textsf{STATE}~\textsf{P1SYMBOL}^\ast~\textsf{HEAD}
\end{align*}%
}%
The unary query $\textsf{LastConf}'[x]$ uses the variable at the beginning of the expression as its answer.
It is easy to verify that the elements accepted by $P'_1$ but not by $P_2$ encode sequences of quasi-configurations of $\mathcal{M}'$ in space $s'$ with respect to these queries. To apply Lemma~\ref{lemma_atmquasienctoenc}, we need to specify an additional $\textsf{SameCell}'$ query for this encoding.

$\textsf{SameCell}'$ is expressed by an \kmq{k+1} query that can in general not be expressed by a \kmq{k} query:

\noindent{\small%
\begin{align*}
\textsf{FirstCell}(\lambda_1,x) &\to\colorpred_1(x)\\
\colorpred_1(x)\wedge\textsf{NextCell}(x,x') &\to\colorpred_1(x')\\
%
\textsf{State}_q(\lambda_1)\wedge\textsf{FirstCell}(\lambda_1,x)\wedge\textsf{Symbol}(x,z)\wedge\textsf{Head}(x,v)\wedge{}\\[-0.7ex]
\textsf{State}_q(\lambda_2)\wedge\textsf{FirstCell}(\lambda_2,y)\wedge\textsf{Symbol}(y,z)\wedge\textsf{Head}(y,v) &\to\colorpred_2(y)\\[-0.7ex]
	& \phantom{{}\to{}}\text{for all $q\in Q$}\\
\colorpred_1(x)\wedge\colorpred_2(y)\wedge\textsf{SameCell}(x,y)\wedge{}\\[-0.7ex]
\textsf{NextCell}(x,x')\wedge\textsf{Symbol}(x',z)\wedge\textsf{Head}(x',v)\wedge{}\\[-0.7ex]
\textsf{NextCell}(y,y')\wedge\textsf{Symbol}(y',z)\wedge\textsf{Head}(y',v) &\to\colorpred_2(y')\\
\colorpred_2(y)\wedge \textsf{LastCell}(y)&\to\checkpred
%
%
\end{align*}}%
where $\textsf{FirstCell}$, $\textsf{Symbol}$, $\textsf{SameCell}$, and $\textsf{LastCell}$ are the queries
for which $P_1$ and $P_2$ containment-encode runs of $\mathcal{M}$.
Note that our constructions already ensure that the sequences of $\mathcal{M}$-cells compared by $\textsf{SameCell}'$
are of the same length.

To complete the proof, we apply Lemma~\ref{lemma_atmquasienctoenc} to construct an \kmq{k+1} $\bar{P}_2$.
The \kmq{k+1} $P_2'$ is obtained by expressing the disjunction of $P_2$ and $\bar{P}_2$ as an \kmq{k+1}
using Proposition~\ref{prop_posfcqs}.
Then $P_1'$ and $P_2'$ containment encode accepting runs of $\mathcal{M}'$ in space $s'$.
\end{proof}

\begin{reptheorem}{theo_hardmdlmqk}
\containmentHardnessStatement{\mdatalogconst}{\kmq{k}}{\kExpTime{$(k+2)$}}
\end{reptheorem}
\begin{proof}
The claim is shown by induction on $k$.
For the base case, we show that deciding containment of $\mq$ queries is \ThreeExpTime-hard.
By Lemma~\ref{lemma_mducqexpatm}, for any DTM $\mathcal{M}^0$, there is a
\mdatalogconst query $P^0_1$, a \linmq $P^0_2$, \linmq{}s as in Definition~\ref{def_atmencode},
and a same-cell query that is a \ucq with respect to which $P^0_1$ and $P^0_2$ containment-encode accepting runs of
$\mathcal{M}^0$ in exponential space $s$.
By applying Lemma~\ref{lemma_atmencodenesting}, we obtain, for an arbitrary ATM $\mathcal{M}^1$,
a \mdatalogconst query $P^1_1$, an \mq $P^1_2$, and \mq queries as in Definition~\ref{def_atmencode} (including a same-cell query),
that containment-encode accepting runs of $\mathcal{M}^1$ in space $s'\geq 2^s$.

The induction step for $k>1$ is immediate from Lemma~\ref{lemma_atmencodenesting}.
\end{proof}

\end{document}